\documentclass{JHEP3}

\usepackage{amssymb}
\usepackage{amsmath}
\usepackage{epsfig}

\makeatletter
\def\simgt{\mathrel{\lower2.5pt\vbox{\lineskip=0pt\baselineskip=0pt
           \hbox{$>$}\hbox{$\sim$}}}}
\def\simlt{\mathrel{\lower2.5pt\vbox{\lineskip=0pt\baselineskip=0pt
           \hbox{$<$}\hbox{$\sim$}}}}
\makeatother

\newcommand{\be}{\begin{equation}}
\newcommand{\ee}{\end{equation}}
\newcommand{\bea}{\begin{eqnarray}}
\newcommand{\eea}{\end{eqnarray}}
\DeclareRobustCommand{\Eq}[1]{Eq.~(\ref{#1})}
\DeclareRobustCommand{\Eqs}[2]{Eqs.~(\ref{#1}) and (\ref{#2})}
\DeclareRobustCommand{\Sec}[1]{Sec.~\ref{#1}}

\DeclareRobustCommand{\App}[1]{App.~\ref{#1}}
\DeclareRobustCommand{\Fig}[1]{Fig.~\ref{#1}}
\DeclareRobustCommand{\Figs}[2]{Figs.~\ref{#1} and \ref{#2}}
\DeclareRobustCommand{\Ref}[1]{Ref.~\cite{#1}}

\DeclareGraphicsExtensions{.png,.pdf}

\newcommand{\vev}[1]{\langle #1 \rangle}

\newcommand{\GeV}{\text{GeV}}
\newcommand{\Br}{\text{Br}}

\newcommand{\MPl}{M_{\rm Pl}}
\newcommand{\sslash}[1]{\ensuremath\raisebox{-0.00cm}{{\small\slash}}\hspace{-0.21cm}#1\/}

\title{Goldstini Can Give the Higgs a Boost}

\author{Jesse Thaler and Zachary Thomas \\
Center for Theoretical Physics, Massachusetts Institute of Technology,\\
$\qquad$ Cambridge, MA 02139, USA \vspace{0.05in} \\
E-mail: \email{jthaler@jthaler.net}, \email{ztt@mit.edu}  
}

\preprint{MIT-CTP 4226}

\abstract{Supersymmetric collider phenomenology depends crucially on whether the lightest observable-sector supersymmetric particle (LOSP) decays, and if so, what the LOSP decay products are.  For instance, in SUSY models where the gravitino is lighter than the LOSP, the LOSP decays to its superpartner and a longitudinal gravitino via supercurrent couplings.  In this paper, we show that LOSP decays can be substantially modified when there are multiple sectors that break supersymmetry, where in addition to the gravitino there are light uneaten goldstini.  As a particularly striking example, a bino-like LOSP can have a near 100\% branching fraction to a higgs boson and an uneaten goldstino, even if the LOSP has negligible higgsino fraction.  This occurs because the uneaten goldstino is unconstrained by the supercurrent, allowing additional operators to mediate LOSP decay.   These operators can be enhanced in the presence of a $U(1)_R$ symmetry, leading to copious boosted higgs production in SUSY cascade decays.
}

\begin{document}

\section{Introduction}

Supersymmetry (SUSY) is a well-motivated extension of the standard model (SM) with rich phenomenological implications for collider experiments like the LHC.    
Most SUSY theories consist of an ``observable sector'' coupled to one or more ``hidden sectors."  The observable sector contains the fields of the supersymmetric standard model (SSM), in particular the lightest observable-sector supersymmetric particle (LOSP).  The hidden sectors are responsible for breaking SUSY and generating soft masses for SM superpartners, and may contain light states accessible to colliders.   

A typical SUSY collider event involves production of two heavy SM superpartners which then undergo cascade decays to a pair of LOSPs.  If there are hidden sector particles lighter than the LOSP, then the subsequent LOSP decays---if they occur inside the detector---can dramatically impact SUSY collider phenomenology.  The most well-known example of a decaying LOSP is when the light hidden sector particle is a gravitino \cite{Dimopoulos:1996vz,Ambrosanio:1996zr,Dimopoulos:1996va,Ambrosanio:1996jn}.  In that case, the LOSP decays to its superpartner and a longitudinal gravitino via interactions constrained by the conserved supercurrent and the goldstino equivalence theorem \cite{Fayet:1977vd,Fayet:1979yb,Casalbuoni:1988kv,Casalbuoni:1988qd}.  For example, a mostly bino LOSP will decay to a photon, $Z$, or---through its small higgsino fraction---a higgs boson.

\FIGURE[t]{
$\qquad$$\qquad$$\qquad$$\qquad$ \includegraphics[scale=0.25]{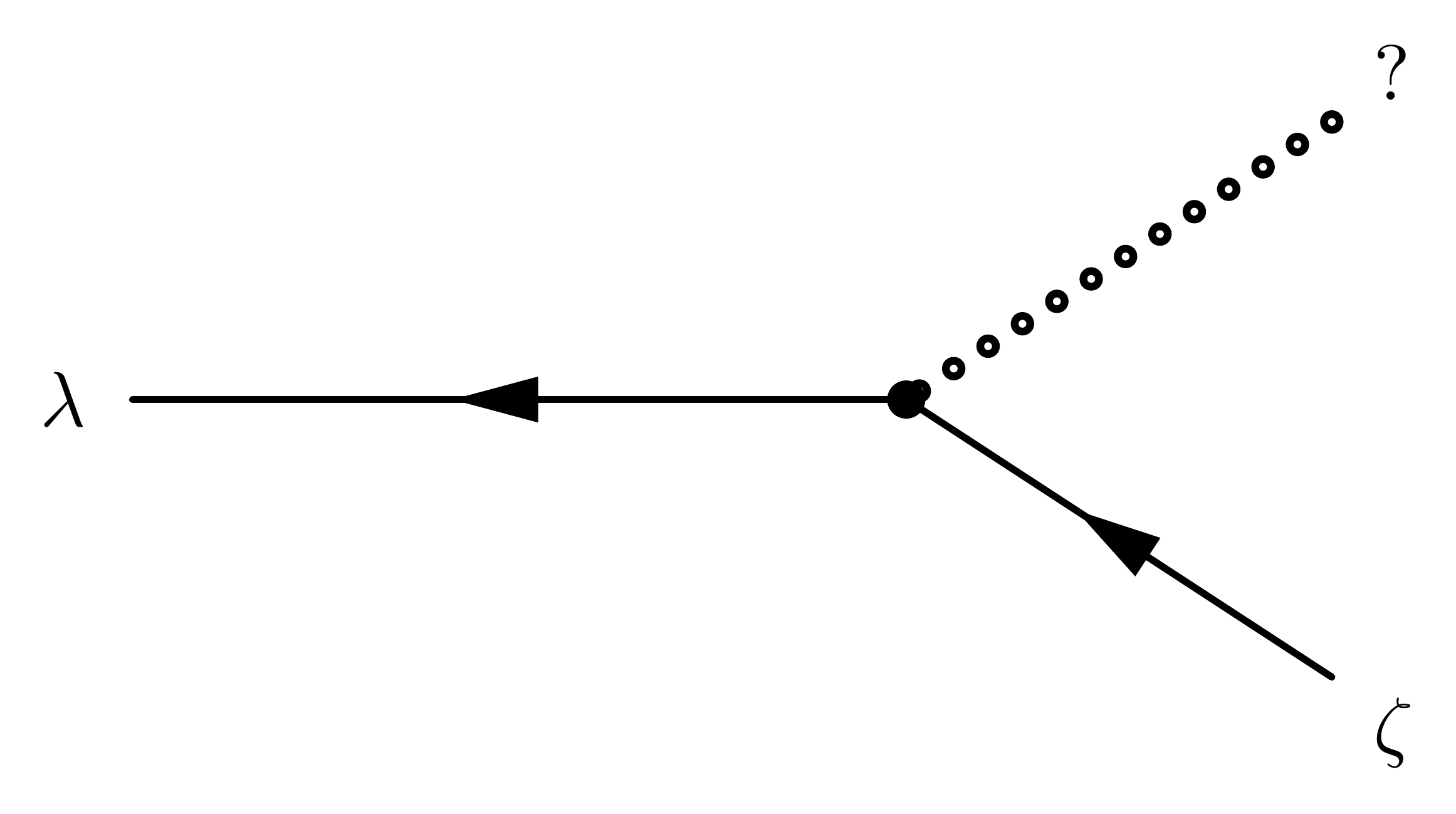}$\qquad$$\qquad$$\qquad$$\qquad$
\caption{A generic LOSP decay.  We will focus on the case where $\lambda$ is a bino-like LOSP, and $\zeta$ is a (pseudo-)goldstino from spontaneous SUSY breaking.  Contrary to the naive expectation, $\lambda$ can decay dominantly to higgs bosons, even if $\lambda$ has negligible higgsino fraction.}
\label{fig:LOSPdecay}
}

In this paper, we will show how changes in the couplings between the observable and hidden sectors can have a dramatic impact on the decay modes of the LOSP, shown  generically in \Fig{fig:LOSPdecay}.  Our case study will be a nearly pure bino LOSP $\lambda$ with an order one branching fraction to higgs bosons, a very counterintuitive decay pattern from the point of view of the standard decay of a bino LOSP to a $\gamma/Z$ plus a longitudinal gravitino.  In fact, in this example, the LOSP branching ratio to higgs bosons is \emph{enhanced} with increasing higgsino mass $\mu$, approaching 100\% in the small $(m_\lambda \tan \beta) / \mu$ limit.  This is unlike the case of a higgsino LOSP, which generically has equal branching fractions to higgs and $Z$ bosons.

These novel bino LOSP decays are possible in the presence of multiple sectors which break supersymmetry, yielding a corresponding multiplicity of ``goldstini'' \cite{Cheung:2010mc}.  While the couplings of the true goldstino (eaten by the gravitino) are constrained by the supercurrent, the orthogonal uneaten goldstini can have different couplings from the naive expectation.  The spectrum of goldstini exhibits a number of fascinating properties \cite{Cheung:2010mc,Craig:2010yf,Argurio:2011hs}, and they may play a role in cosmology or dark matter \cite{Cheung:2010qf,Cheng:2010mw}.  Here, we will focus on properties of goldstini relevant for their collider phenomenology.

\FIGURE[t]{
$\qquad$ \includegraphics[scale=0.6]{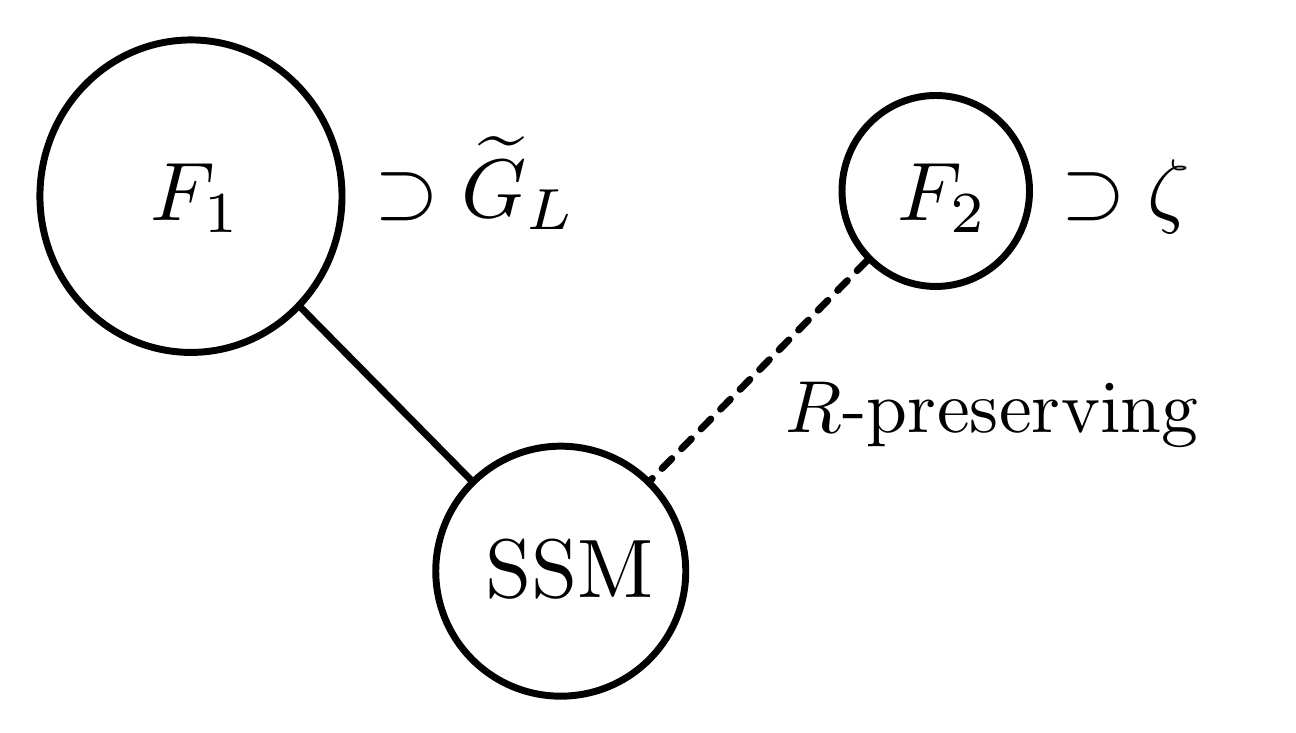} $\qquad$ 
\caption{The $R$-symmetric setup that will be the focus of this paper.  Here, sector 1 has a higher SUSY breaking scale than sector 2, i.e.\ $F_1 \gg F_2$, so the LOSP preferentially decays to the pseudo-goldstino $\zeta$ coming mostly from sector 2.  Since sector 2 preserves an $R$ symmetry, the decay $\lambda \rightarrow \gamma/Z + \zeta$ is highly suppressed, and the mode $\lambda \rightarrow h^0 + \zeta$ can dominate.}
\label{fig:Rsetup}
}

For our case study, we consider two sectors which break SUSY, both of which communicate to the SSM, but one of which preserves an $U(1)_R$ symmetry, as in \Fig{fig:Rsetup}.\footnote{There have been recent studies where the entire SUSY breaking and SSM sectors preserve a $U(1)_R$ symmetry \cite{Kribs:2007ac,Amigo:2008rc}.}  For the appropriate hierarchy of SUSY breaking scales, the LOSP will couple more strongly to the uneaten goldstino $\zeta$ than to the longitudinal gravitino $\widetilde{G}_L$.  Since the uneaten goldstino $\zeta$ is charged under the $U(1)_R$ symmetry, the $R$-violating decay $\lambda \rightarrow \gamma/Z + \zeta$ is suppressed, letting the counterintuitive decay $\lambda \rightarrow h^0 + \zeta $ dominate.\footnote{In \Ref{Cheung:2010mc}, it was erroneously claimed that in the presence of an $R$ symmetry, the dominant decay is $\lambda \rightarrow \psi \bar{\psi} + \zeta $, where $\psi$ is a SM fermion.  This paper corrects that error.}  This fascinating result is demonstrated in \Fig{fig:xiplot}.

\FIGURE[t]{
$\qquad$ \includegraphics[scale=0.8]{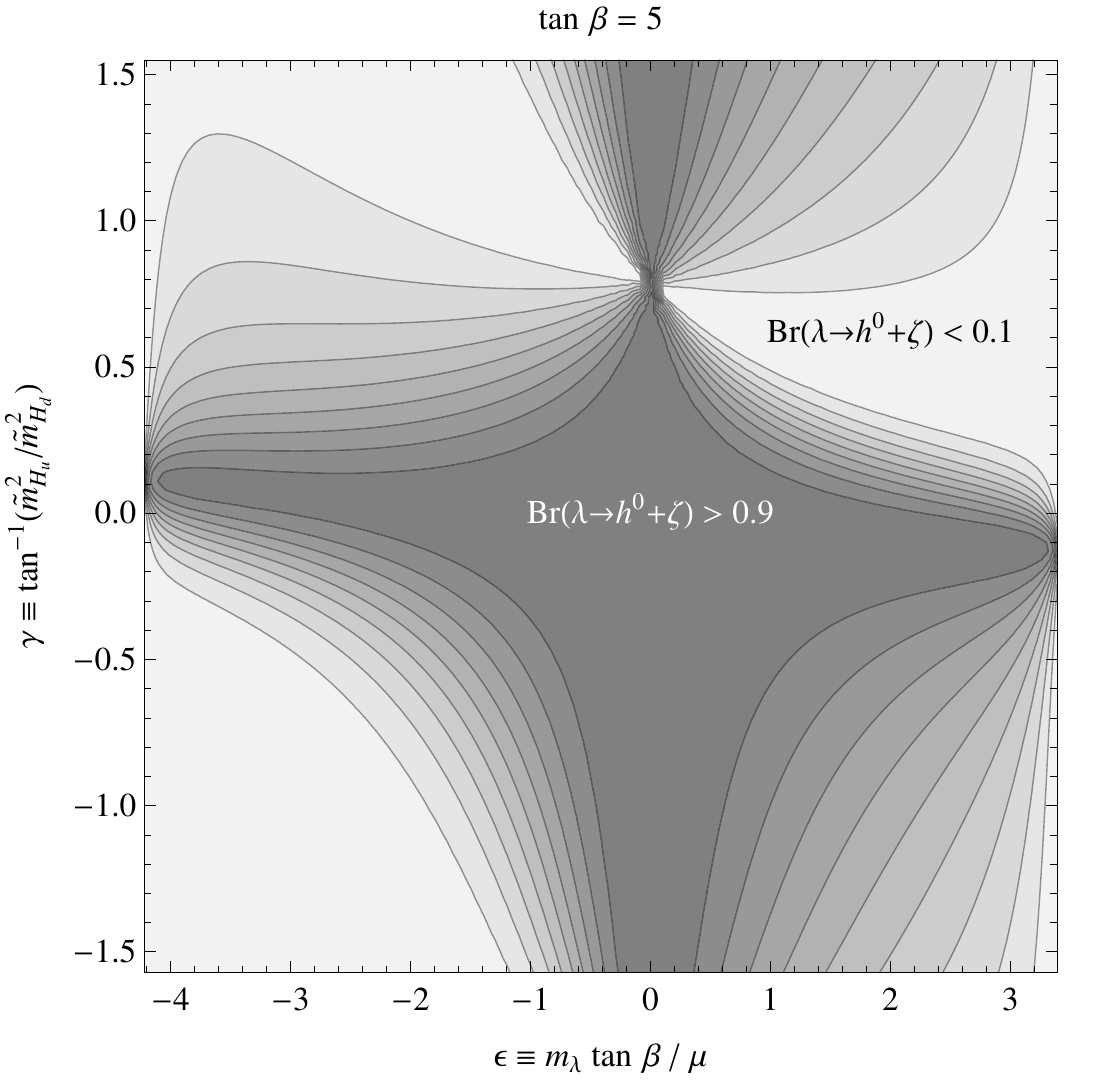} $\qquad$
\caption{Branching ratio $\lambda \to h^0 + \zeta$ for a bino LOSP in the $R$-symmetric setup from \Fig{fig:Rsetup}.  Throughout this parameter space, the remaining branching ratio is dominated by $\lambda \to Z + \zeta$.  The expected mode $\lambda \to \gamma + \zeta$ is almost entirely absent.  Shown is $\Br(\lambda \to h^0 \zeta)$ as a function of $\epsilon \equiv m_\lambda \tan \beta / \mu$ and $\gamma \equiv \tan^{-1} (\widetilde{m}^2_{H_u}/\widetilde{m}^2_{H_d})$, fixing $\tan \beta = 5$, $M_1 = 155~\GeV$, and $m_{h^0} = 120~\GeV$.  The plot terminates on the left and right side at the kinematic bound $m_\lambda < m_{h^0}$.}
\label{fig:xiplot}
}

In this way, goldstini can give the higgs a boost:  a boost in production cross section since most LOSP decays yield a higgs boson; and a boost in kinematics since the higgses are produced with relatively large gamma factors in SUSY cascade decays.  This example gives further motivation to identify boosted higgses using jet substructure techniques \cite{Butterworth:2008iy,Kribs:2009yh,Abdesselam:2010pt}.  This example also motivates searches for other counter-intuitive LOSP decay patterns, where there is a mismatch between the identity of the LOSP and its decay products.

In the next section, we summarize and explain the main results of this paper.  We then describe the framework of goldstini in \Sec{sec:framework}, and derive the low energy effective goldstini interactions and resulting LOSP decay widths in \Sec{sec:eft}.  We explain in more detail why the goldstini case differs from the more familiar gravitino case in \Sec{sec:gravitinocase}.  Plots of the LOSP branching ratios appear in \Sec{sec:results}, and we conclude in \Sec{sec:conclude}.  Various calculational details are left to the appendices.

\section{Counterintuitive LOSP Decays}
\label{sec-executive}

Throughout this paper, we will be considering the situation where a LOSP decays to a lighter neutral fermion as in \Fig{fig:LOSPdecay}, and we will assume the minimal SSM (MSSM) field content.  The possible decay patterns of a LOSP are constrained by symmetries, at minimum conservation of SM charges.  In the familiar case where the LOSP decays to its superpartner and a gravitino, there are further constraints imposed by conservation of the supercurrent.  We will see that these constraints can be significantly relaxed in the presence of multiple SUSY breaking sectors. 

\subsection{A Conventional Goldstino}

In the conventional setup with a single SUSY breaking sector and a light gravitino, the couplings of the helicity-1/2 components of the gravitino are linked via the goldstino equivalence theorem to the couplings of the goldstino $\widetilde{G}_L$.  Supercurrent conservation implies that, at leading order in the inverse SUSY breaking scale $1/F$, the goldstino couples only derivatively to observable sector fields via the supercurrent:
\begin{eqnarray}
\mathcal{L}_{\textrm{eff}} & = & i \widetilde{G}_L^\dagger \bar{\sigma}^\mu \partial_\mu \widetilde{G}_L + \frac{1}{F} \partial_\mu \widetilde{G}_L j^\mu \label{eq:Leff-supercurrent}, \\
j^\mu & = & \sigma^\nu \bar{\sigma}^\mu \psi_i D_\nu \phi^{* i} - \frac{1}{2 \sqrt{2}} \sigma^\nu \bar{\sigma}^\rho \sigma^\mu \lambda^{\dagger a} F_{\nu \rho}^a, \label{eq:supercurrent}
\end{eqnarray}
where we have elided terms that vanish on the goldstino equation of motion.  Here, $\phi_i$ is a scalar and $\psi_i$ is its fermionic superpartner, and $F^a_{\mu \nu}$ is a gauge field strength with $\lambda^a$ its corresponding gaugino.  In particular, the only possible LOSP decays are to its superpartner and a gravitino.  This implies, for example, that a pure right-handed stau LOSP $\tilde{\tau}_R$ can only decay to a gravitino and a right-helicity tau $\tau_R$, despite the fact that after electroweak symmetry breaking, there is no symmetry forbidding the decay to a left-helicity tau $\tau_L$.

\FIGURE[t]{
\includegraphics[scale=0.25]{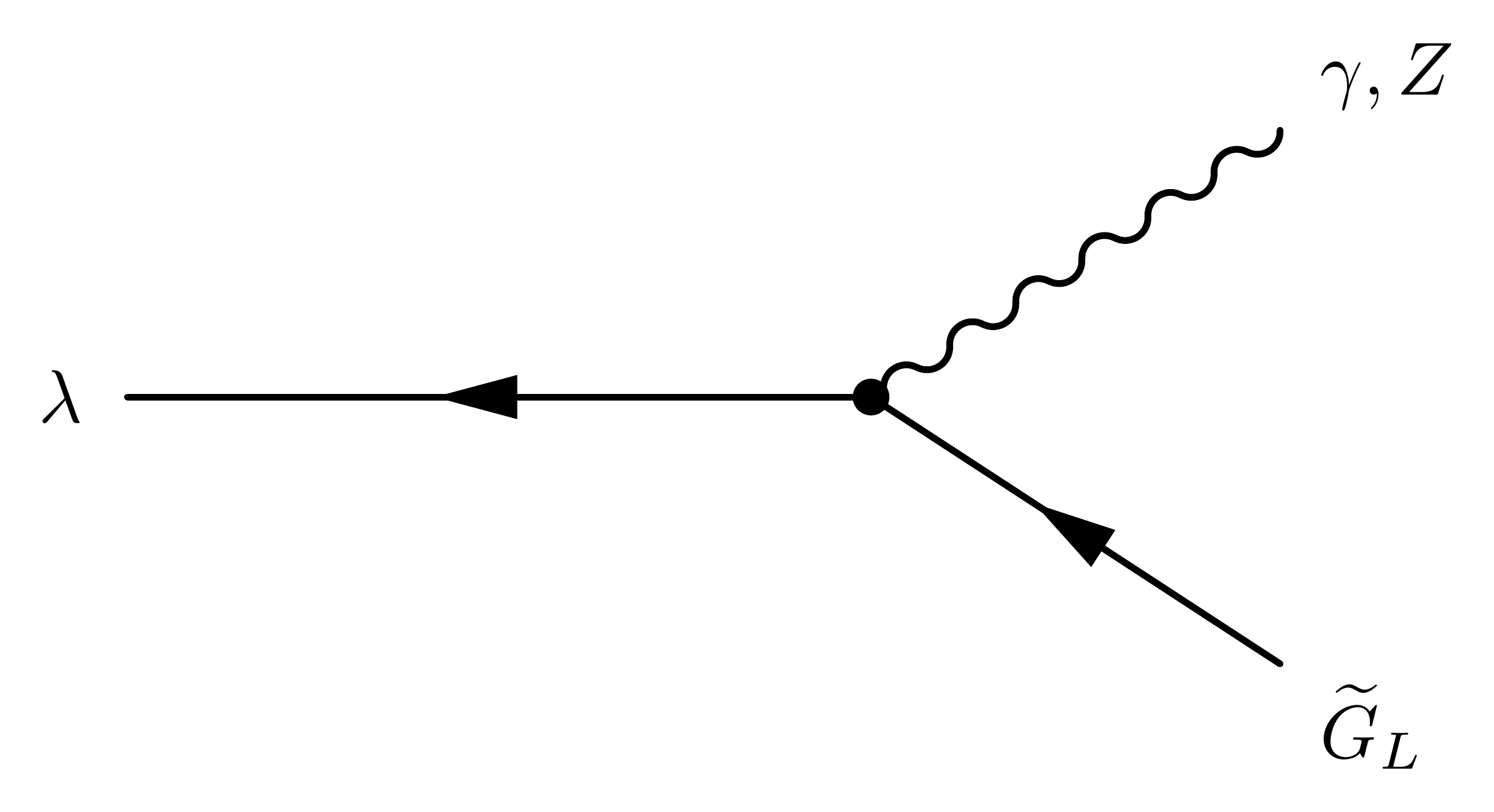}
$\qquad$
\includegraphics[scale=0.25]{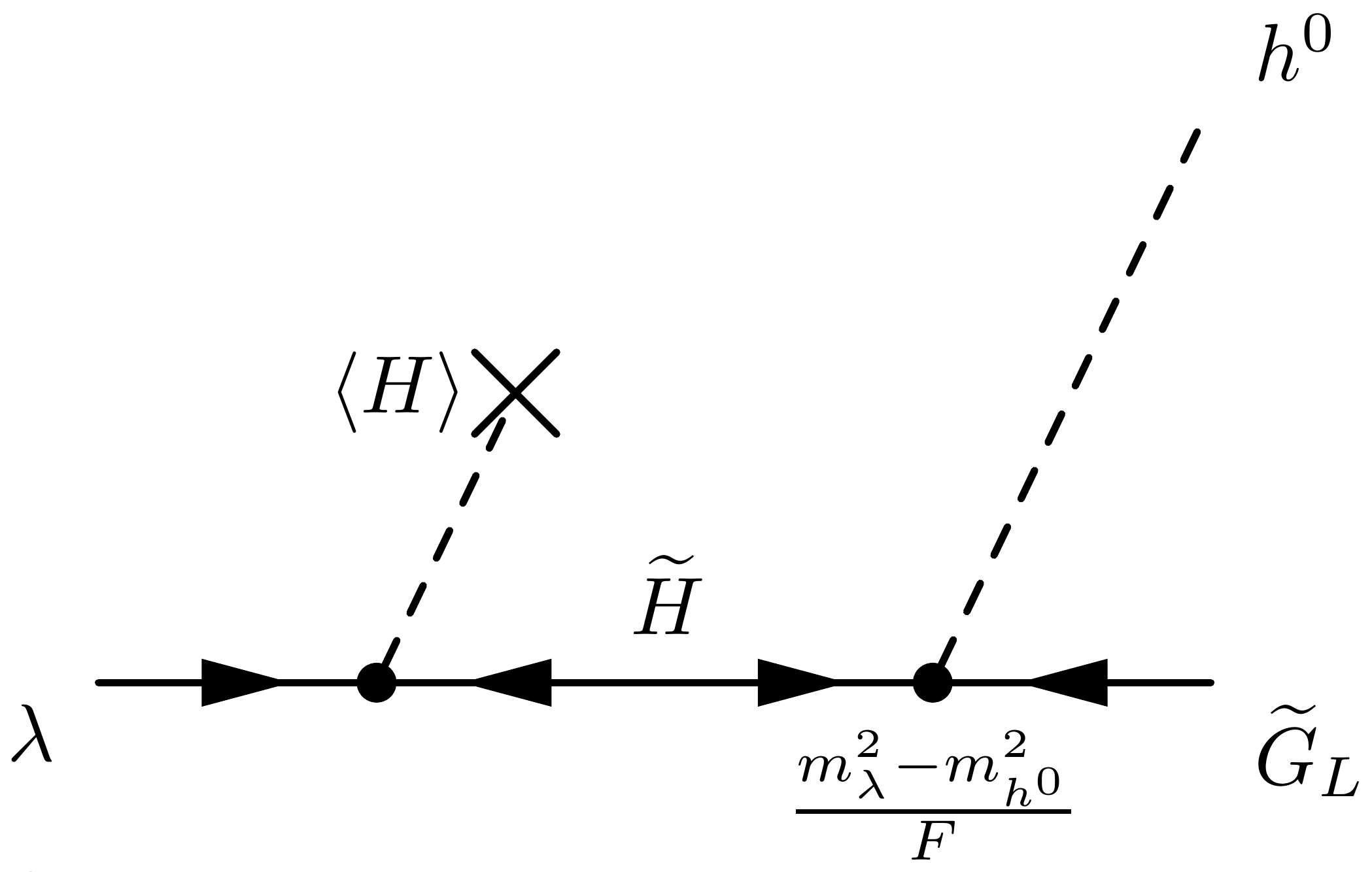}
\caption{The standard decays of a bino-like LOSP to the longitudinal gravitino.  They are primarily to a photon or $Z$ (left), though the bino may also decay to a higgs via its higgsino component (right).  The derivatives in \Eq{eq:supercurrent} yield the Yukawa coupling labeled here, proportional to the mass-squared difference of the on-shell bino and higgs.  A cancellation between the two possible intermediate higgsinos means the propagator contributes a factor of $\mu^{-2}$ to the amplitude at leading order, leading to a very large suppression of this channel in the higgsino decoupling limit.  Feynman diagrams throughout follow the conventions of \Ref{Dreiner:2008tw}.}
\label{fig:standarddecays}
}

For concreteness we will focus on a bino-like LOSP throughout this paper, though many of the following arguments hold with only minor modifications for a wino, as well.  In that case, the supercurrent in \Eq{eq:supercurrent} permits the decay $\lambda \rightarrow \gamma/Z + \widetilde{G}_L$ via the second term in the supercurrent.  There is also a possible decay $\lambda \rightarrow h^0 + \widetilde{G}_L$ where $h^0$ is the physical higgs boson, but since this occurs entirely through the higgsino fraction of the LOSP, it will be comparatively suppressed.\footnote{See \Ref{Meade:2009qv} for a recent discussion of more general neutralino decays.}  Explicitly, to leading order in $m_\lambda / \mu$, the dominant LOSP partial widths are
\begin{eqnarray}
\Gamma_\gamma & = & \frac{m_\lambda^5 \cos^2 \theta_W}{16 \pi F^2}, \label{eq:gravgamma}\\
\Gamma_Z & = & \frac{m_\lambda^5 \sin^2 \theta_W}{16 \pi F^2} \left(1 - \frac{M_Z^2}{m_\lambda^2}\right)^4, \label{eq:gravZ}
\eea
where $m_\lambda \simeq M_1$ is the bino-like LOSP mass, and $\theta_W$ is the weak mixing angle.  The subdominant width to higgs bosons is
\bea
\Gamma_{h^0} & = & \frac{m_\lambda^2 M_Z^2}{\mu^4} \frac{m_\lambda^5 \sin^2 \theta_W \cos^2 2 \beta}{32 \pi F^2} \left(1 - \frac{m_{h^0}^2}{m_\lambda^2}\right)^2, \label{eq:gravh}
\eea
where $\tan \beta \equiv v_u/v_d$.  Feynman diagrams for these standard decays are shown in \Fig{fig:standarddecays}.

\subsection{Additional Operators?}

In the case of the true goldstino $\widetilde{G}_L$, its couplings are saturated by \Eq{eq:supercurrent}.  But if the LOSP were to decay not to a true goldstino but to a generic neutral fermion $\zeta$, then there are many more operators that might mediate LOSP decay instead.  For example, the dimension 5 operator
\begin{align}
\mathcal{O}^5_R& =  C^5_R \frac{\mu}{F} \lambda \zeta (H_u \cdot H_d)^* &  (\lambda \rightarrow h^0 +  \zeta) \label{eq:op5} 
\end{align}
mediates the decay  $\lambda \rightarrow h^0 +  \zeta$ after electroweak symmetry breaking.  Here, the coefficient $\mu/F$ has been chosen with malice aforethought, as this will turn out to be the approximate scaling behavior for the \emph{eaten} goldstino. The subscript $R$ indicates that this operator will preserve a $U(1)_R$ symmetry once we identify $\zeta$ with an uneaten goldstino of $R$-charge 1.  

There are also additional operators at dimension 5 which violate this $U(1)_R$ symmetry,   
\bea
\mathcal{O}^5_{\sslash{R},u \cdot d} & = &  C^5_{\sslash{R},u\cdot d} \frac{ \mu}{F} \lambda \zeta (H_u \cdot H_d),\\
\mathcal{O}^5_{\sslash{R},u} & = &  C^5_{\sslash{R},u} \frac{ \mu}{F} \lambda \zeta H_u^\dagger H_u, \\
\mathcal{O}^5_{\sslash{R},d} & = &  C^5_{\sslash{R},d} \frac{ \mu}{F}  \lambda \zeta H_d^\dagger H_d.
\eea
Considering these $\mathcal{O}^5$ operators together, the partial width for the decay $\lambda \rightarrow h^0 + \zeta$ is
\be
\Gamma_{h^0}  = \left(C^5_{\textrm{net}}\right)^2 \frac{\mu^2 M_Z^2}{m_\lambda^4} \frac{m_\lambda^5 \sin^2 \theta_W}{32 \pi F^2} \left(1 - \frac{m_{h^0}^2}{m_\lambda^2}\right)^2.
\ee
Here, we have defined
\be
C^5_{\textrm{net}}  =  \frac{\sqrt{2}}{g'} \left( \left(C^5_R + C^5_{\sslash{R},u\cdot d}\right) \cos (\alpha + \beta) - 2 C^5_{\sslash{R},u} \sin \beta \cos \alpha + 2 C^5_{\sslash{R},d} \cos \beta \sin \alpha \right), \label{eq:y}
\ee
with $\alpha$ being the physical higgs mixing angle.  Thus, if somehow the $\mathcal{O}^5$ operators were dominant over operators like those in \Eq{eq:supercurrent}, then the decay of a pure bino LOSP to a higgs would dominate over the decay to a $\gamma/Z$.  Note that the $\mathcal{O}^5$ operators only mediate a decay to one or more higgs bosons, and not to a longitudinal $Z$, due to the gauge invariance of the scalar portion of the operators.  

\FIGURE[t]{
\includegraphics[scale=0.25]{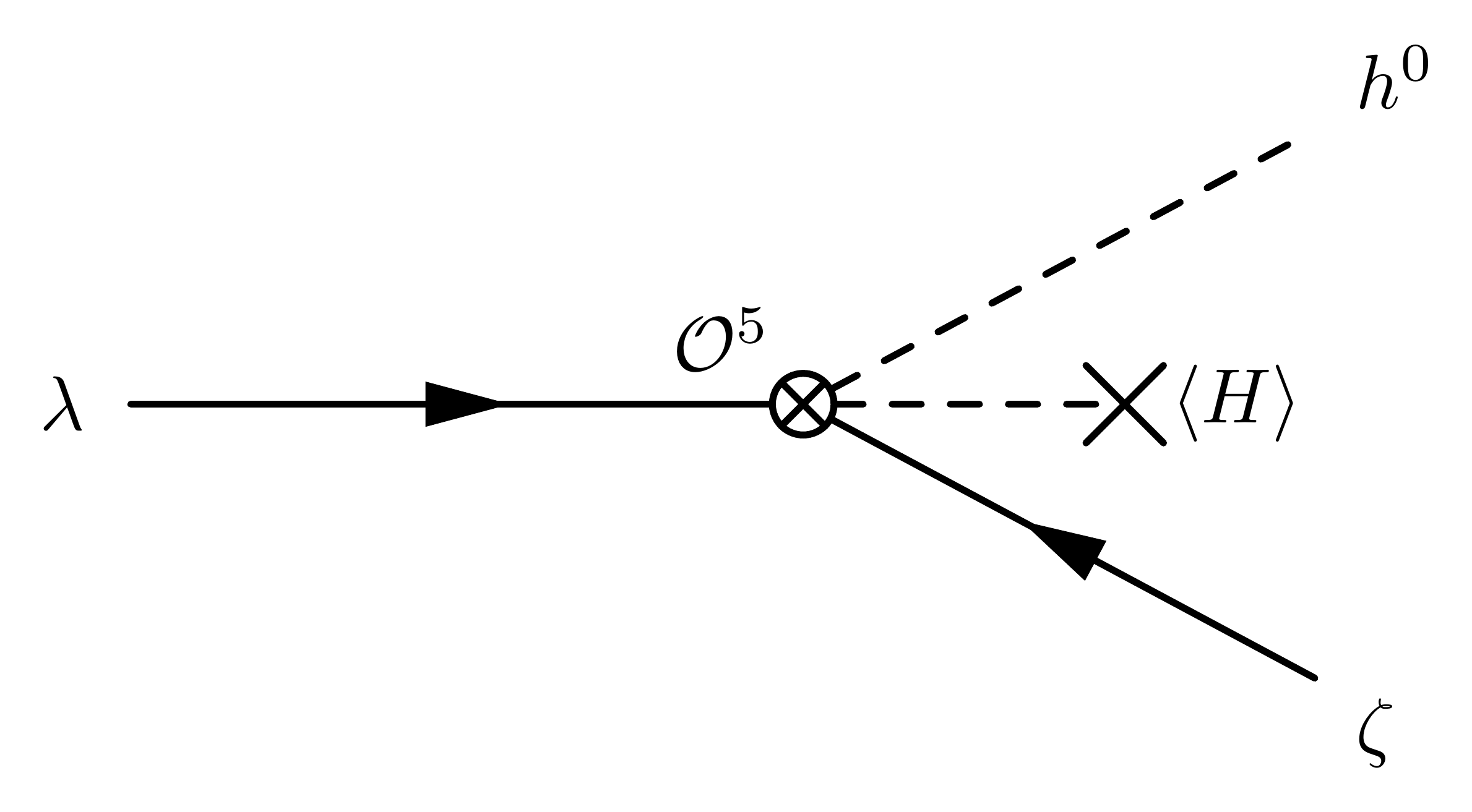}
$\qquad$
\includegraphics[scale=0.25]{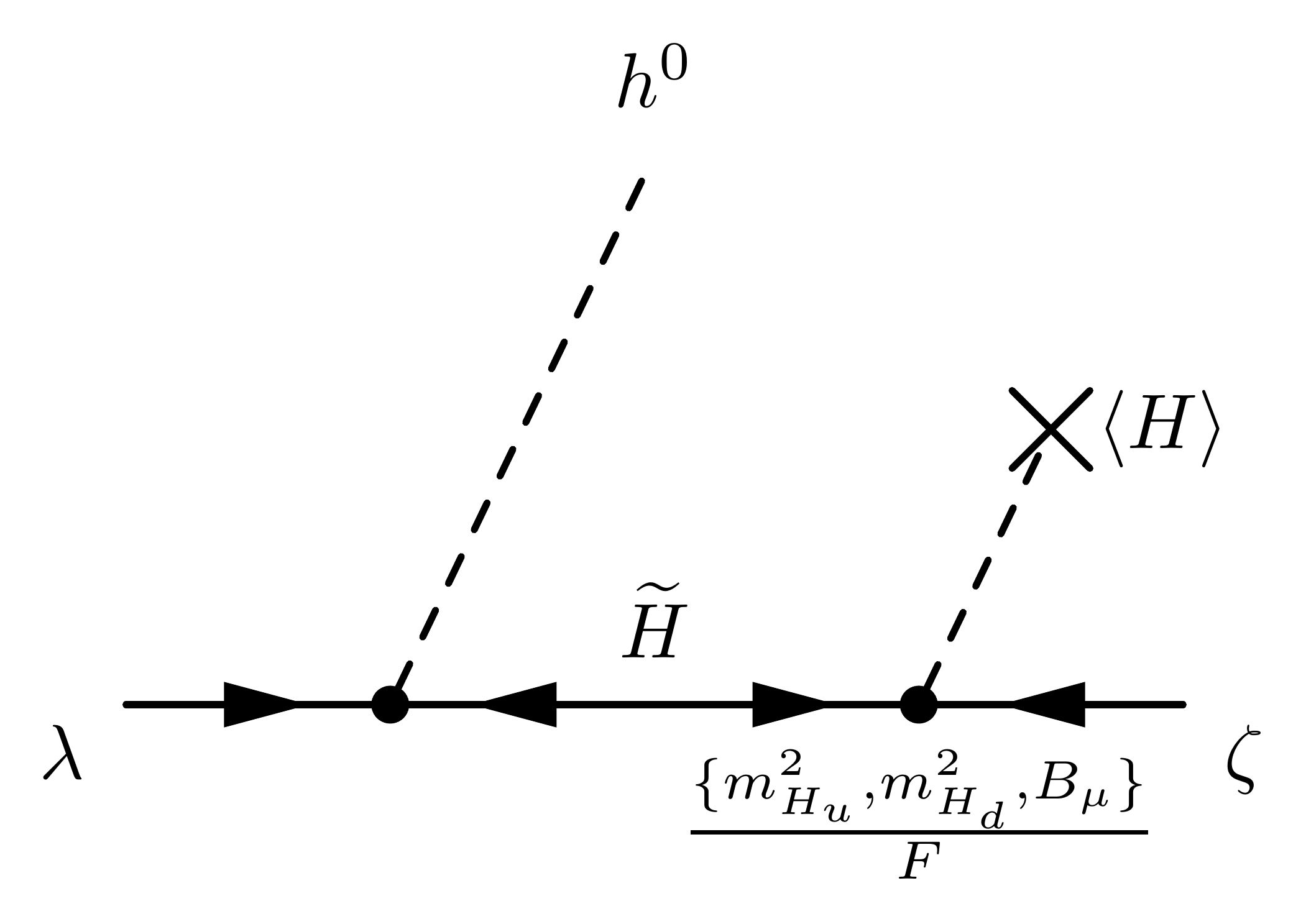}
\caption{Additional diagrams which could contribute to LOSP decay.  The dimension 5 operator (left) can be generated by integrating out an intermediate higgsino (right).  There is also a diagram with $h^0$ and $\vev{H}$ reversed.  However, if $\zeta$ is a longitudinal gravitino $\widetilde{G}_L$, then the width $\Gamma(\lambda \rightarrow h^0 + \zeta)$ vanishes in the higgsino decoupling limit.}
\label{fig:dim5decay}
}

Now, in the conventional goldstino case, there is a sense in which the $\mathcal{O}^5$ operators are indeed generated after integrating out the higgsino as in  \Fig{fig:dim5decay}.  This occurs not in the derivatively-coupled basis, but rather in the non-linear goldstino basis described in \Sec{sec:framework}.  The pertinent combination of Wilson coefficients attains the value 
\be
C^5_{\textrm{net}} = \frac{(m_{H_u}^2 - m_{H_d}^2) \sin 2 \beta + 2 B_\mu \cos 2 \beta}{\mu^2} + \mathcal{O}\left(\frac{m_\lambda}{\mu}\right), \label{eq:c5net-2}
\ee
which MSSM aficionados will recognize as being \emph{zero} for the tree-level higgs potential in the decoupling limit $|\mu| \gg M_Z$---the same limit in which it was legitimate to integrate out the higgsinos in the first place (see \App{app:treehiggs} for an explanation of this cancellation).  This is as it must be; the physical predictions in this field basis must agree with those of the basis corresponding to the supercurrent picture of \Eq{eq:supercurrent}, in which the decay rate to higgs bosons is highly suppressed.

However, because $C^5_{\textrm{net}} = 0$ arises only because of a delicate cancellation in the true goldstino case, any deviation will give rise to additional LOSP decays beyond the supercurrent prediction. In particular, if there are multiple sectors that break SUSY \cite{Cheung:2010mc}, each of which contributes only partially to the SSM soft masses, then the couplings of the uneaten goldstini cannot be determined by supercurrent considerations.\footnote{This fact was recently exploited in \Ref{Cheng:2010mw} to arrange for goldstini dark matter with leptophilic decays.}   In general, the goldstini will have very different couplings from the gravitino; concretely, the goldstini need not be derivatively coupled to observable-sector particles.  For a generic uneaten goldstino
\be
C^5_{\textrm{net}} = \frac{(\widetilde{m}_{H_u}^2 - \widetilde{m}_{H_d}^2) \sin 2 \beta + 2 \widetilde{B}_\mu \cos 2 \beta}{\mu^2} + \mathcal{O}\left(\frac{m_\lambda}{\mu}\right),
\ee
where the tildes indicate the linear combination, appropriate to the given goldstino, of contributions from the SUSY-breaking sectors to the corresponding soft mass.   These parameters need not cancel and thus a pure bino LOSP can exhibit the counterintuitive decay to a higgs boson and an uneaten goldstino.

\subsection{Goldstini and $R$ Symmetries}
\label{sec:goldstiniandR}

The differences between LOSP decays to an eaten goldstino versus an uneaten goldstino become especially striking in the presence of a $U(1)_R$ symmetry, and they will be the main example in this paper.  Consider the case of two SUSY breaking sectors  as in \Fig{fig:Rsetup} where the uneaten goldstino is associated with a sector 2 that preserves an $R$-symmetry.  As we will argue in \Sec{sec:framework}, if the scale of SUSY breaking in sector 1 is much higher than in sector 2, i.e. $F_1 \gg F_2$, then we can ignore the standard LOSP decay to a gravitino, since it will be overwhelmed by the LOSP decay to the uneaten goldstino from sector 2. 

\FIGURE[t]{
\includegraphics[scale=0.20]{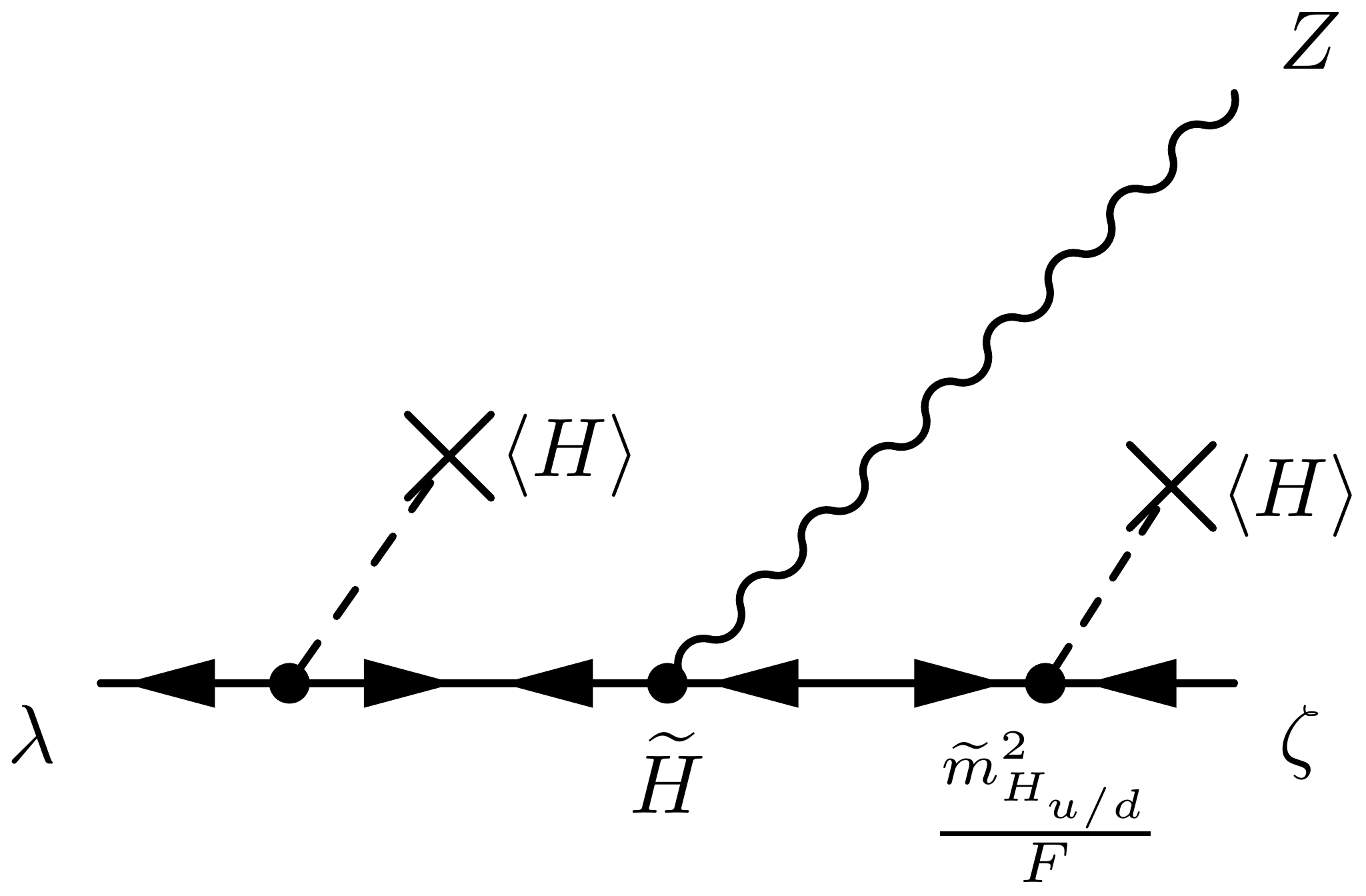} $\qquad$
\includegraphics[scale=0.20]{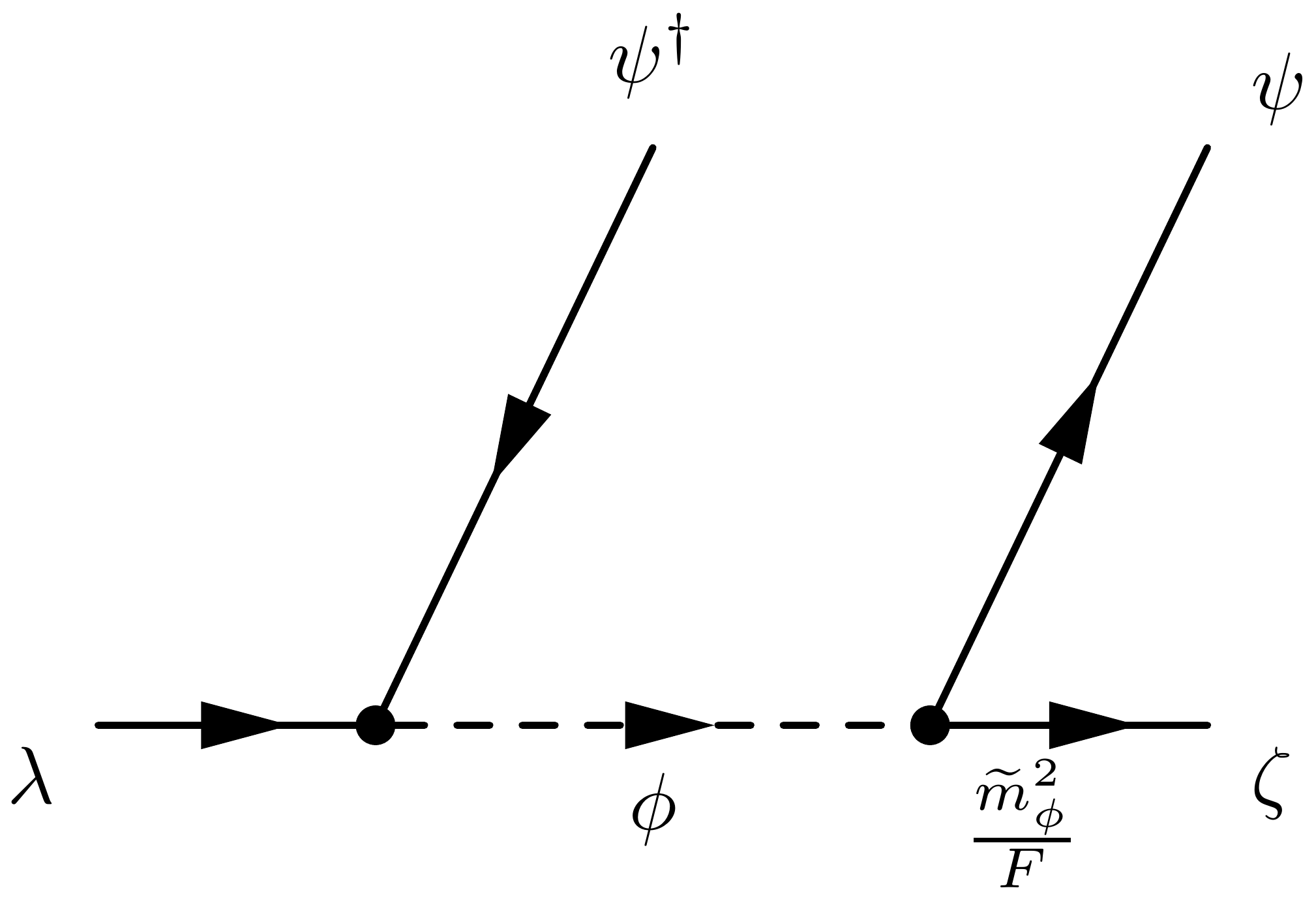}$\qquad$
\includegraphics[scale=0.20]{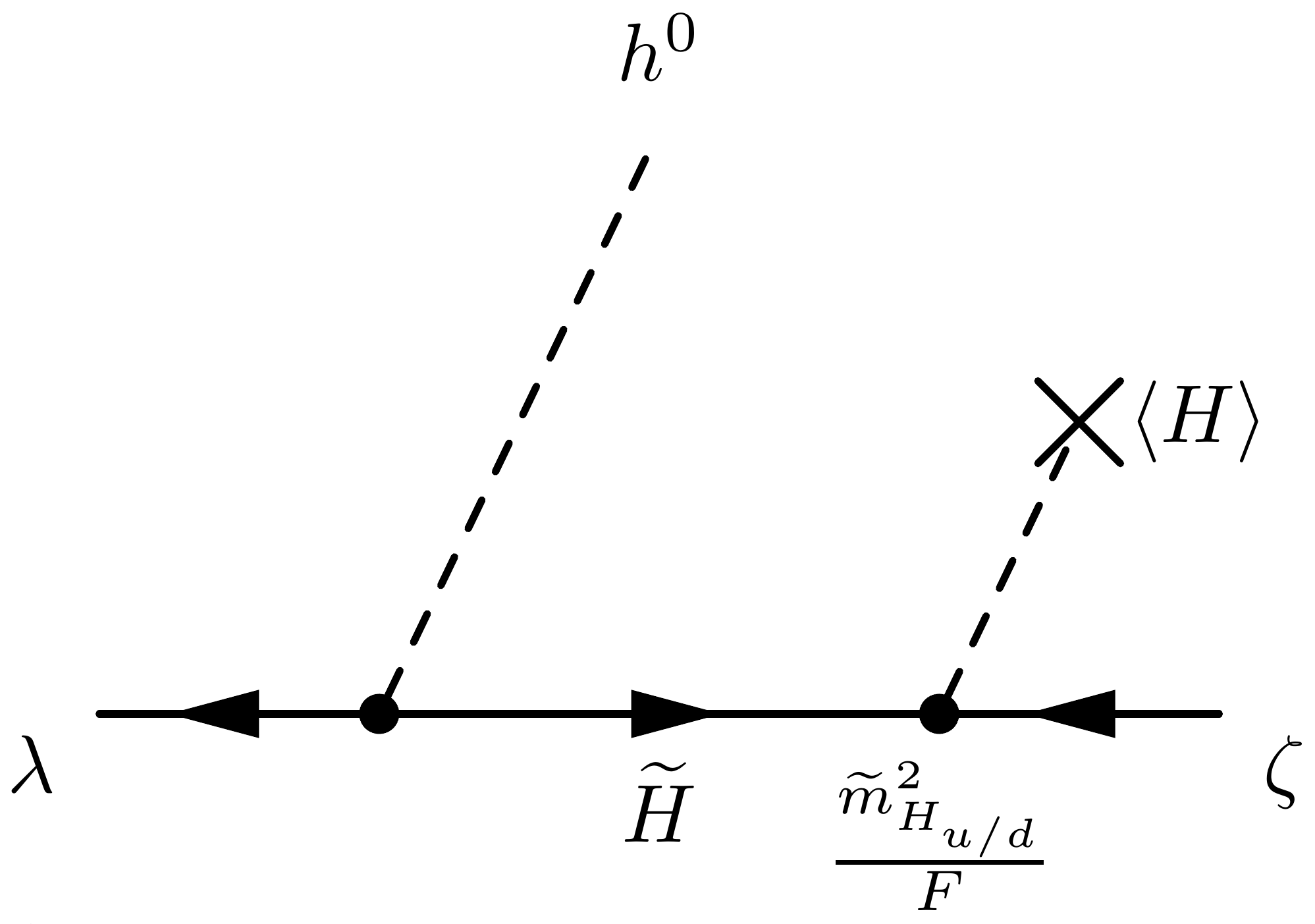}

\caption{Representative diagrams contributing to the dimension 6 operators.  After integrating out the intermediate higgsinos or sfermions, these diagrams mediate LOSP decays to $Z$ bosons and SM difermions, as well as generating additional LOSP decays to $h^0$.}
\label{fig:Rdecays}
}

The gaugino soft mass terms violate the $R$-symmetry, so a bino LOSP cannot undergo the associated decay to a $\gamma$/$Z$ and the uneaten goldstino $\zeta$.  Instead, it \emph{must} (at tree level) decay to the uneaten goldstino via a virtual higgsino or sfermion as in \Figs{fig:dim5decay}{fig:Rdecays}, producing a higgs $h^0$, an arbitarily-polarized $Z$, or two SM fermions $\psi\bar{\psi}$ in the process.

To understand this effect more clearly, note there are only a limited number of $R$-symmetric operators that can mediate the decay of a bino LOSP to an uneaten goldstino and standard model particles once the higgsinos and sfermions are integrated out.  At dimension 5, only $\mathcal{O}^5_R$ respects the $R$-symmetry; the other $\mathcal{O}^5$ operators are associated with the $R$-symmetry-violating $B_\mu$ term.  At dimension 6,  we will show that the only operators consistent with gauge symmetries, $R$-parity, and our imposed $R$-symmetry are
\begin{align}
\mathcal{O}^6_{\Phi,1} & = \frac{C^6_{\Phi,1}}{F} i \zeta^\dagger \bar{\sigma}^\mu \lambda \Phi^\dagger D_\mu \Phi & (\lambda \rightarrow h^0/Z +  \zeta), \label{eq:op6-1} \\
\mathcal{O}^6_{\Phi,2} & = \frac{C^6_{\Phi,2}}{F} i \zeta^\dagger \bar{\sigma}^\mu \lambda (D_\mu \Phi^\dagger) \Phi  & (\lambda \rightarrow h^0/Z +  \zeta), \label{eq:op6-2}\\
\mathcal{O}^6_\psi & = \frac{C^6_\psi}{F} (\zeta^\dagger \psi^\dagger) (\psi \lambda) & (\lambda \rightarrow \psi \bar{\psi} + \zeta), \label{eq:op6-psi}
\end{align}
where $\Phi$ stands for either $H_u$ or $H_d$, $\psi$ is an SM fermion, and we have indicated in parentheses the corresponding LOSP decay mode.  The values of the Wilson coefficients $C^6$ are omitted here for clarity; they are given explicitly in \Eq{eq:WilsonForR}.  Despite the fact that we have integrated out a higgsino/sfermion, these operators are not suppressed by the higgsino/sfermion mass as there is a cancellation between the propagator of the virtual heavy particle and its coupling to the goldstino.  We will explain this fact in more detail in \Sec{sec:eft}; it is sufficient to note for now that the $\mathcal{O}^6_\Phi$ are suppressed by a power of $\mu$ relative to $\mathcal{O}^5_R$.

The relative importance of $\mathcal{O}^5_R$, $\mathcal{O}^6_{\Phi,i}$, and $\mathcal{O}^6_\psi$ for LOSP decays depend sensitively on the SSM parameters.  In general, the three-body decay $\lambda \rightarrow \psi \bar{\psi} + \zeta$ is subdominant to the two-body decays $\lambda \rightarrow h^0/Z +  \zeta$.  As mentioned already, $\mathcal{O}^5_R$ only mediates a decay to higgs bosons, not to longitudinal $Z$ bosons, whereas $\mathcal{O}^6_{\Phi,i}$ can yield either, or even a transverse $Z$.  One might naively expect $\mathcal{O}^5_R$ to dominate over $\mathcal{O}^6_{\Phi,i}$, since the dimension 6 operator has a decay amplitude suppressed by $m_\lambda/\mu$.  However, $\mathcal{O}^5_R$ contains $H_u \cdot H_d$ which involves an additional  $1 / \tan \beta$ suppression in the large $\tan \beta$ limit, while the operators $\mathcal{O}^6_{H_u,i}$ have no such suppression.   Thus, the dimension 6 decays are only suppressed by
\be
\label{eq:xidef}
\epsilon \equiv \frac{m_\lambda \tan \beta}{\mu}
\ee
compared to the dimension 5 decays, which may not even be a suppression at large $\tan \beta$.  

In \Fig{fig:xiplot}, we showed the LOSP branching ratios as a function of both $\epsilon$ and the most important other free parameter in the theory
\be
\label{eq:gammadef}
\tan \gamma \equiv \frac{\widetilde{m}^2_{H_u}}{\widetilde{m}^2_{H_d}},
\ee
which is the ratio of the contributions to $m_{H_u}^2$ and $m_{H_d}^2$ from the sector containing the uneaten goldstino.  For special values of $\gamma$, the decay mode $\lambda \to Z + \zeta$ can either be completely suppressed or enhanced relative to $\lambda \to h^0 + \zeta$ due to cancellations.  Our main interest will be in the higgsino decoupling limit with small $\epsilon$, where the higgs mode generically dominates.  

Thus, in the presence of a $R$-symmetry, the LOSP decay to an uneaten goldstino gives a boost to higgs boson production, even if (and especially if) the LOSP has a negligible higgsino fraction.  Moreover, the decays to the uneaten goldstino, whether featuring a higgs boson or not, will completely dominate over any decays to the gravitino if there is an appropriate hierarchy between the two SUSY-breaking scales, as we will describe in the next section.

\section{Goldstino and Gravitino Couplings}
\label{sec:framework}

Having understood the possibility of enhanced $\lambda \rightarrow h^0 +  \zeta$ decays from an operator perspective, the remainder of this paper will show how precisely this works in the explicit example of multiple SUSY breaking sectors.

\subsection{The General Framework}

As in \Ref{Cheung:2010mc}, we consider two sequestered sectors, each of which spontaneously breaks SUSY.    Each sector has an associated goldstino ($\eta_1$ and $\eta_2$, respectively), and we characterize the size of SUSY breaking via the goldstino decay constants ($F_1$ and $F_2$, respectively).  Each SUSY breaking sector can be parametrized in terms of a non-linear goldstino multiplet \cite{Komargodski:2009rz,Cheung:2010mc}
\be
X_i = \frac{\eta_i^2}{2F_i} + \sqrt{2} \theta \eta_i + \theta^2 F_i,
\ee
for $i = 1, 2$.  We define the quantities
\be
F \equiv \sqrt{F_1^2 + F_2^2}, \qquad \tan \theta \equiv \frac{F_2}{F_1},
\ee
and we take $\tan \theta \le 1$ ($F_1 \ge F_2$) without loss of generality.  

The combination $\widetilde{G}_L = \sin \theta \, \eta_1 + \cos \theta \, \eta_2$ is eaten by the gravitino to become its longitudinal components via the super-higgs mechanism, but the orthogonal goldstino $\zeta = \cos \theta \, \eta_1 - \sin \theta \, \eta_2$ remains uneaten and will be the focus of our study.    For simplicity, we will work in the $\MPl \rightarrow \infty$ limit where the uneaten goldstino remains massless, though in general $\zeta$ will get a mass proportional to $m_{3/2}$ via SUGRA effects, in particular $m_\zeta = 2m_{3/2}$ in the minimal goldstini scenario \cite{Cheung:2010mc}.  In addition, variations in the SUSY-breaking dynamics \cite{Craig:2010yf} or induced couplings between the two sectors \cite{Cheung:2010mc,Argurio:2011hs} can modify the mass term for $\zeta$.\footnote{At minimum, one expects loops of SM fields to generate $m_\zeta \simeq m_{\rm soft}/(16\pi^2)^n$ \cite{Cheung:2010mc}, where $n$ depends on the number of loops necessary to effectively connect sectors 1 and 2 and transmit the needed $U(1)_R$ breaking.  The uneaten goldstino will also obtain a tree-level mass due to mixing with the neutralinos, but this is of order $1/F^2$ and is comparatively negligible.}

Supersymmetry breaking is communicated from the two hidden sectors to the visible sector by means of a non-trivial K\"{a}hler potential and gauge kinetic function (presumably coming from integrating out heavy messenger fields).  Some representative terms contributing to the SSM soft masses are\footnote{We only give the K\"{a}hler potential for a single species of scalar; more general $A$ and $B$-terms involving multiple species can also be formed.}
\begin{eqnarray}
K & = & \Phi^\dagger \Phi \sum_i \frac{m^2_{\phi,i}}{F_i} X_i^\dagger X_i, \\
f_{a b} & = & \frac{1}{g_a^2} \delta_{a b}\left(1 +\sum_i \frac{2 M_{a,i}}{F_i} X_i\right),
\end{eqnarray}
where $i = 1,2$, and $\Phi$ stands for a general SSM multiplet.  These yield the following terms in the lagrangian up to order $1/F$ \cite{Binetruy:2000zx}:
\begin{eqnarray}
\mathcal{L} & = & - \sum_i m^2_{\phi,i} \phi^* \phi + \sum_i \frac{m^2_{\phi,i}}{F_i} \eta_i \psi \phi^* \nonumber \\
& &  - \frac{1}{2} \sum_i M_{a,i} \lambda^a \lambda^a  - \sum_i \frac{i M_{a,i}}{\sqrt{2} F_i} \eta_i \sigma^{\mu \nu} \lambda^a  F_{\mu \nu}^a  + \sum_i \frac{M_{a,i}}{\sqrt{2} F_i} \eta_i \lambda^a D^a .
\label{eq:fulllagrangian}
\end{eqnarray}
Thus, the parameter $m_{\phi,i}^2$ ($M_{a,i}$) is the contribution to the SUSY-breaking scalar (gaugino) mass from each respective sector.  Note that they are intrinsically related to the coupling of the SSM fields to the goldstini.  

Rotating to the $\widetilde{G}_L$--$\zeta$ basis yields similar interaction terms for the eaten goldstino $\widetilde{G}_L$ and the uneaten goldstino $\zeta$,
\begin{eqnarray}
\mathcal{L}_{\widetilde{G}_L} & = & \frac{m_\phi^2}{F} \widetilde{G}_L \psi \phi^* - \frac{i M_a}{\sqrt{2} F} \widetilde{G}_L \sigma^{\mu \nu} \lambda^a F^a_{\mu \nu}+ \frac{M_a}{\sqrt{2} F} \widetilde{G}_L \lambda^a D^a, \label{Lgrav} \\
\mathcal{L}_{\zeta} & = & \frac{\widetilde{m}_\phi^2}{F} \zeta \psi \phi^* - \frac{i \widetilde{M}_a}{\sqrt{2} F} \zeta \sigma^{\mu \nu} \lambda^a F^a_{\mu \nu} + \frac{\widetilde{M}_a}{\sqrt{2} F} \zeta \lambda^a D^a, \label{Lgold}
\end{eqnarray}
where the untilded and tilded mass parameters associated with gauginos denote
\begin{eqnarray}
M_a& = & M_{a,1} + M_{a,2},\\
\widetilde{M}_a & = & M_{a,2} \cot \theta - M_{a,1} \tan \theta, \label{eq:tilde}
\end{eqnarray}
with the analogous notation for the scalar mass-squared parameters.  Throughout, we will work in the limit $\cot \theta \gg 1$, for which we can take
\be
\frac{\widetilde{M}_a}{F} = \frac{M_{a,2}}{F_2}, \qquad \frac{\widetilde{m}^2_\phi}{F} = \frac{m^2_{\phi,2}}{F_2}.
\ee
In this limit, as long as any of the $M_{a,2}$ or $m_{\phi,2}^2$ are at least on the order of the weak scale, LOSP decays to gravitinos are very suppressed and can be ignored for collider purposes.  We see that as predicted via the supercurrent, the true goldstino $\widetilde{G}_L$ couples to SSM fields in proportion to the physical soft masses.  In contrast, $\zeta$ couples via the \emph{tilded} mass parameters which in the $\cot \theta \gg 1$ limit are proportional just to the contribution of sector 2 to the SSM soft masses.

\subsection{The Decoupling and $R$-symmetric Limit}
\label{sec:decouplingandRlimit}

In this paper, we will focus on the higgsino decoupling and $R$-symmetric limits.  That is, we will be considering the limit where $\mu$ is large compared to $m_\lambda$, and the limit where sector 2 preserves a $U(1)_R$ symmetry.  There are a number of important features of this limit.  

When the higgsinos are decoupled, the soft terms $m_{H_u}^2$, $m_{H_d}^2$, and $B_\mu$ must scale as $\mathcal{O}(\mu^2)$ in order to get successful electroweak symmetry breaking.\footnote{Strictly speaking, this is only true for the combinations $m_{H_u}^2$ and $m_{H_d}^2 + B_\mu \tan \beta$ (working in the large $\tan \beta$ limit).  However, if one simultaneously decouples the heavy higgs scalars in the same way, so that $m_{A^0}^2$ is of order $\mu^2$, then all three soft mass parameters scale as $\mu^2$ barring accidental cancellations.  Our later results for the uneaten goldstino are robust against this assumption, since $m_{H_u}^2$ has the desired scaling properties regardless.}  We can see from the above lagrangian that the coupling of $\widetilde{G}_L$ to a higgsino and a higgs is proportional to these $\mathcal{O}(\mu^2)$ soft SUSY-breaking masses.  The same is true for the couplings of $\zeta$ if we make the additional simplifying assumption that the \emph{tilded} mass parameters scale in the same fashion, so long as this is not forbidden by a symmetry.  With one noted exception in \Sec{sec:difermions}, however, our results do not depend on this assumption.  From the diagrams in \Fig{fig:dim5decay}, one would naively expect the amplitudes for the decay of a bino LOSP to the physical higgs and either goldstino via a virtual higgsino to be of order $\mu$ and thus dominant over other decays to the same goldstino in the decoupling limit.  As we will argue in \Sec{subsec:cancellations}, there is a cancellation in the $\widetilde{G}_L$ case which renders the decay $\lambda \rightarrow h^0 +  \widetilde{G}_L$ small, whereas for $\zeta$, the decay $\lambda \rightarrow h^0 +  \zeta$ can indeed dominate.

In the limit where sector 2 is $R$-symmetric, the contribution from sector 2 to SSM $A$-terms, $B$-terms, and gaugino masses is zero.  Most relevant for our purposes, this implies that $\widetilde{B}_\mu$ and $\widetilde{M}_1$ are nearly zero.  The absence of a $\widetilde{B}_\mu$ term implies that the cancellation in \Eq{eq:c5net-2} seen for $\widetilde{G}_L$ cannot persist for the uneaten goldstino $\zeta$.  The absence of a $\widetilde{M}_1$ term means that the LOSP decay to a $\gamma/Z$ and $\zeta$ is highly suppressed.\footnote{In the alternative limit where sector 1 preserves an $R$-symmetry, one expects $\lambda \rightarrow \gamma/Z + \zeta$ to still be relevant, but that will not be the focus of this paper.}  Both of these facts imply a large $\lambda \rightarrow h^0 +  \zeta$ branching fraction.  Depending on the relative importance of the dimension 5 or dimension 6 operators, the mode $\lambda \rightarrow Z +  \zeta$ can be large as well.

\section{Higgsino Decoupling Limit Effective Field Theory}
\label{sec:eft}

Starting from the above goldstini framework, we can now systematically describe which operators contribute to bino LOSP decay in the higgsino decoupling and $R$-symmetric limits.  We will then give the resulting decay rates for the three main decay modes: $\lambda \rightarrow h^0 + \zeta$, $\lambda \rightarrow Z + \zeta$, and $\lambda \rightarrow \psi \bar{\psi} + \zeta$.

\subsection{Leading $R$-symmetric Operators}
\label{subsec:RsymOps}

In the higgsino decoupling limit, it is convenient to organize the LOSP decay operators in terms of the small parameter $m_\lambda/\mu$.  This may be accomplished practically by integrating out the heavy higgsino degrees of freedom, yielding an effective field theory with successively higher-dimension operators suppressed by additional powers of $\mu$.  Away from the decoupling limit, \App{sec:allorders} describes how to calculate the LOSP branching fractions for arbitrary $\mu$.   For simplicity, we will take $F_1 \gg F_2$, in which case the couplings of the uneaten goldstino are completely determined by sector 2.  

Recall that in the MSSM, gauginos have $R$-charge 1, higgs multiplets have $R$-charge 1, and matter multiplets have $R$-charge $1/2$.  For an $R$-symmetric SUSY breaking sector, the corresponding goldstino has $R$-charge 1.  Putting this together, at dimension 5, there is only a single operator contributing to bino LOSP decay consistent with the symmetries of the theory (including the imposed $R$-symmetry):  
\be
\mathcal{O}^5_R = C^5_R \frac{\mu}{F} \lambda \zeta (H_u \cdot H_d)^*. 
\ee
This operator may mediate the decay of a bino LOSP to the uneaten goldstino and one or two physical higgs bosons $h^0$.\footnote{Gauge invariance of the scalar portion of the operator forbids production of goldstone bosons (i.e.~longitudinal $W/Z$ bosons), and the heavier higgs bosons $A^0$, $H^0$, and $H^\pm$ are of course kinematically excluded in the decoupling limit.}  

At dimension 6, there are three sorts of additional operators:\footnote{We have used integration by parts to move all derivatives off of $\lambda$, and used field redefinitions to eliminate terms proportional to the equations of motion of the goldstino and gauge bosons.  We elect not to use field redefinitions to eliminate terms proportional to $\bar{\sigma}^\mu \partial_\mu \lambda$, as the resulting operators (arising from the gaugino mass term) would violate the $R$-symmetry.}
\bea
\mathcal{O}^6_{\Phi,1} & =&  \frac{C^6_{\Phi,1}}{F} i \zeta^\dagger \bar{\sigma}^\mu \lambda \Phi^\dagger D_\mu \Phi,  \\
\mathcal{O}^6_{\Phi,2} & = & \frac{C^6_{\Phi,2}}{F} i \zeta^\dagger \bar{\sigma}^\mu \lambda (D_\mu \Phi^\dagger) \Phi,  \\
\mathcal{O}^6_\psi & = & \frac{C^6_\psi}{F} (\zeta^\dagger \psi^\dagger) (\psi \lambda),
\eea
where $\Phi$ stands for either $H_u$ or $H_d$, and $\psi$ is a standard model fermion.   The dimension 6 operators $\mathcal{O}^6_{\Phi,i}$ may produce a $Z$ boson (longitudinal or otherwise) instead of or in addition to any higgs boson production.  The dimension 6 operator $\mathcal{O}^6_\psi$ will produce a difermion pair instead.\footnote{This operator arises from integrating out intermediate sfermions as opposed to higgsinos, so our power counting may be spoiled if there are any relatively light sfermions.  We will later explicitly calculate the decay rate for $\lambda \rightarrow \bar{\psi}\psi + \zeta$ at tree level to all orders in $m_\lambda^2/m_\phi^2$ to account for this possibility.}  The effects of $\mathcal{O}^6_\psi$, but not the others, were considered in \Ref{Cheung:2010mc}.

We have omitted two possible $R$-symmetric operators, $\partial^\mu \zeta \sigma^\nu \lambda^\dagger F_{\mu \nu}$ and $\partial^\mu \zeta \sigma^\nu \lambda^\dagger \widetilde{F}_{\mu \nu}$, which could mediate the decay of the bino to a photon or $Z$ and the goldstino.  It is clear by examining the original lagrangian of \Eq{Lgold} that in the $R$-symmetric limit with $\widetilde{M}_1 = 0$, a decay to a photon cannot occur at tree-level, so that any effects of such operators will be suppressed compared to the others of the same mass dimension.

The values of the Wilson coefficients for the above operators can be found by matching onto the original lagrangian of \Eq{Lgold}:
\begin{align}
C^5_R & = \frac{g' \left(\widetilde{m}_{H_u}^2 - \widetilde{m}_{H_d}^2\right)}{\mu^2}, & C^6_{H_u,1} & =  \frac{g' \widetilde{m}_{H_u}^2 }{\sqrt{2} \mu^2}, & C^6_{H_d,1} & =  - \frac{g' \widetilde{m}_{H_d}^2}{\sqrt{2} \mu^2},\nonumber \\
C^6_\psi & =  - \sqrt{2} g' Y_\psi \frac{\widetilde{m}^2_\phi}{m_\phi^2}, & C^6_{H_u,2} & = 0, & C^6_{H_d,2} & = 0.  \label{eq:WilsonForR}
\end{align}
Here, $g'$ is the hypercharge gauge coupling, $Y_\psi$ is the hypercharge of the relevant SM fermion, and the tilded mass parameters are defined in \Eq{eq:tilde}.  Inverse powers of the higgsino mass-squared $\mu^2$ and scalar mass-squared $m_\phi^2$ appear as expected, since these are the masses of the fields we are integrating out. 

The key observation is that the above Wilson coefficients are still order $\mathcal{O}(\mu^0)$ in the higgsino decoupling limit,\footnote{Note that they are not of order 1, but rather of order $\cot \theta$.  We have chosen to leave such dependence in the Wilson coefficients, rather than replacing $F$ with $F_2$ everywhere, so that the only modification needed to describe the couplings of the \emph{eaten} goldstino is to remove tildes from all soft mass parameters.} since the soft masses scale as  $\mathcal{O}(\mu^2)$.  Thus, even if the LOSP has negligible higgsino fraction, there are relevant bino-goldstino-higgs couplings.  As advertised, the leading decays in the higgsino-decoupling and $R$-symmetric limits are
\be
\lambda \rightarrow h^0 + \zeta, \qquad \lambda \rightarrow Z + \zeta, \qquad \lambda \rightarrow  \psi \bar{\psi} + \zeta.
\ee
Now, using the effective operators of \Sec{subsec:RsymOps}, we can calculate the various bino LOSP decay widths in the higgsino decoupling and $R$-symmetric limits.  Possible $R$-violating decays are described in \App{app:rviolatingdecays}.

\subsection{Decay to Higgs Bosons}

The contributions to the $\lambda \rightarrow h^0 + \zeta$ decay from the dimension 5 and dimension 6 operators may be expressed in terms of an effective Yukawa interaction for on-shell states:\footnote{This is not strictly speaking the whole story; the bino may also decay via two local dimension 5 operators ($\mathcal{O}^5_R$ and $\lambda \lambda (H_u \cdot H_d)^*$ or their wino equivalents) connected by a virtual wino or bino.   However, their contributions to the decay amplitude are suppressed by $m_\lambda/(\mu \tan \beta)$ compared to that of $O^5_R$ alone, or $1/\tan^2 \beta$ to those of the dimension 6 operators, and can be safely ignored in most limits.}
\be
\mathcal{L}_{\rm eff} = - \frac{M_Z \mu \sin \theta_W}{\sqrt{2} F} \left( C^5_{\textrm{net}} + \frac{ m_\lambda}{\mu} C^6_{\textrm{net}} \right) \lambda \zeta h^0. \label{eq:yukawa} 
\ee
The coefficients $C^5_{\textrm{net}}$ and $C^6_{\textrm{net}}$ are appropriate linear combinations of the Wilson coefficients of the dimension 5 and 6 operators, respectively, and are given explicitly in \App{app:rviolatingdecays}.  In the decoupling and $R$-symmetric limits, they take on the values
\bea
C^5_{\textrm{net}} & = & \frac{(\widetilde{m}_{H_u}^2 - \widetilde{m}_{H_d}^2 ) \sin 2\beta}{\mu^2}, \\
C^6_{\textrm{net}} & = & \frac{  \widetilde{m}_{H_u}^2 \sin^2 \beta - \widetilde{m}_{H_d}^2 \cos^2 \beta }{\mu^2}.
\eea
The decay rate via this channel is
\be
\Gamma  =  \frac{ m_\lambda \mu^2 M_Z^2 \sin^2 \theta_W }{32 \pi  F^2} \left( C^5_{\textrm{net}} + \frac{m_\lambda}{\mu} C^6_{\textrm{net}} \right)^2 \left(1 - \frac{m_{h^0}^2}{m_{\lambda}^2}\right)^2. \label{eq:decayrate-higgs}
\ee

In the extreme decoupling limit, we would expect the $C^5_{\textrm{net}}$ term, arising from the dimension 5 operator, to dominate over the effects of any dimension 6 operators, which are naturally suppressed by a factor of $m_\lambda/\mu$.  However, our power counting may be spoiled for large $\tan \beta$, due to the factor of $\sin 2 \beta \approx 2 / \tan \beta$ in $C^5_{\textrm{net}}$.  In the event that $\tan \beta$ is of the same order as $\mu/m_\lambda$, we cannot neglect the dimension 6 operators.  There are no such complications for the dimension 7 or higher operators, which may be safely ignored in the decoupling limit.

As a side note, there are only a few changes to the above calculation if we consider a wino LOSP.  There are now \emph{two} allowed operators at dimension 5---namely, $\lambda^a \zeta (H_u T^a \cdot  H_d)^*$ and $\lambda^a \zeta (H_u \cdot T^a H_d)^*$---but the results throughout are almost identical, requiring only the replacement $g' \rightarrow -g$ or $\sin \theta_W \rightarrow - \cos \theta_W$ (as the neutral higgsinos have $T^3$ and $Y$ differing only by a sign).  In particular, one can verify that there is no net coupling to the $Z$ boson from the dimension 5 operators,\footnote{Dimension 5 operators can, however, induce a $\lambda^\pm \rightarrow W^\pm \zeta$ decay.  Such decays may well be phenomenologically interesting, as the competing observable-sector decays ($\lambda^\pm \rightarrow l^\pm \nu \lambda_3$, $\lambda^\pm \rightarrow \pi^\pm \lambda_3$, et al.\@ \cite{Chen:1996ap}) can be highly suppressed due to the near-degeneracy of the chargino and wino.  However, such decays are certainly not specific to this $R$-symmetric limit, or even to the multiple goldstino model.} so the neutral wino LOSP decays dominantly to higgs bosons in the small $(m_\lambda \tan \beta)/\mu$ limit.

\subsection{Decay to $Z$ Bosons}

The dimension 5 operator does not contribute to $Z$ decay.  The dimension 6 operators mediate the decay $\lambda \rightarrow Z + \zeta$ due to the presence of covariant derivatives.  Expanding the lagrangian in unitarity gauge, we find a relatively simple coupling to the $Z$ boson:
\be
\mathcal{L} = \frac{M_Z^2 \sin \theta_W}{\sqrt{2} F} C^6_{\textrm{net},Z} \zeta^\dagger \bar{\sigma}^\mu \lambda Z_\mu,
\ee
with $C^6_{\textrm{net}, Z}$ being a different linear combination of the Wilson coefficients of the dimension 6 operators.  The definition of $C^6_{\textrm{net}, Z}$ is given explicitly in \App{app:rviolatingdecays}, and attains the value 
\be
C^6_{\textrm{net},Z} = - \frac{\widetilde{m}_{H_u}^2 \sin^2 \beta + \widetilde{m}_{H_d}^2 \cos^2 \beta}{\mu^2}
\ee
in the decoupling and $R$-symmetric limit. The resulting decay rate is
\be
\label{eq:decayrate-Z}
\Gamma_Z  = \frac{M_Z^2 m_\lambda^3 \sin^2 \theta_W}{32 \pi F^2} \left(C^6_{\textrm{net},Z}\right)^2  \left(1 - \frac{M_Z^2}{m_\lambda^2}\right)^2  \left(1 + 2 \frac{M_Z^2}{m_\lambda^2} \right).
\ee

\subsection{Decay to Difermions}

Finally, the operator $O^6_\psi$ mediates the decay of a bino LOSP to a goldstino and a fermion pair.   The decay rate from just this operator is
\begin{eqnarray}
\Gamma_{\psi \bar{\psi}} & = & \frac{m_\lambda^5 \sec^2 \theta_W}{32 \pi F^2} \frac{\alpha_{\rm EM} Y_\psi^2}{12 \pi} \frac{ \widetilde{m}_\phi^4}{m_\phi^4}
\label{eq:difermiondecay}
\end{eqnarray}
in the limit of vanishing fermion masses.\footnote{We also neglect here possible contributions from interference between diagrams featuring this operator and diagrams in which the fermions originate from an off-shell higgs or $Z$ produced by one of the other dimension 6 operators.}

As argued in \Ref{Cheung:2010mc}, the decay rate is non-zero even in the limit of very large scalar masses.  However, due to the factor of $\alpha_{\textrm{EM}} Y_\psi^2 / (12 \pi)$, the decay rate to fermions is typically subdominant to the higgs and $Z$ modes, even after summing over all possible fermion final states.  One might wonder whether there could be an enhancement at moderate values of the scalar masses.  Calculating the explicit tree-level decay rate for this mode to all orders in the scalar mass (while still working in the higgsino decoupling limit), the result in \Eq{eq:difermiondecay} is multiplied by a function $f[m_\phi^2 / m_\lambda^2]$:
\begin{eqnarray}
f[x] = 6 x^2 \left(-5 + 6 x + 2 (x-1)(3x - 1) \log \left[1 - \frac{1}{x}\right]\right) \label{f} \simeq 1 + \frac{4}{5 x} + \mathcal{O}\left(\frac{1}{x^2}\right). \label{eq:lightscalar}
\end{eqnarray}
This function never grows larger than 6 (at $m_\phi = m_\lambda$), and drops off quite sharply from that value as $m_\phi$ increases.  For example, $m_\phi$ must be less than $1.25 m_\lambda$ for $f$ to be greater than 2.  Thus, the difermion mode is indeed subdominant.  The sole exception occurs when $\widetilde{m}_{H_u}^2$ and $\widetilde{m}_{H_d}^2$ are both close to zero, where the higgs and $Z$ decay modes are suppressed.

\section{Comparisons to the Gravitino Case}
\label{sec:gravitinocase}

Before showing results for bino LOSP branching ratios in the next section, it is instructive to compare the $R$-symmetric goldstino results in \Sec{sec:eft} to the more familiar case of a gravitino.  Indeed, the existence of a bino-goldstino-higgs coupling in the higgsino decoupling limit is quite surprising from the point of view of the more familiar longitudinal gravitino couplings, where it is known that the decay $\lambda \rightarrow h^0 + \widetilde{G}_L$ is highly suppressed.  In this section, we will go to the higgsino decoupling limit and calculate the effective interactions for a longitudinal gravitino.  In the decoupling limit effective theory, we will find seemingly miraculous cancellations enforced by supercurrent conservation.  

\subsection{Additional Operations for the Gravitino}
\label{subsec:additionaloperators}

In the higgsino decoupling limit for a longitudinal gravitino, the operators from \Sec{subsec:RsymOps} persist after the replacement $\zeta \rightarrow \widetilde{G}_L$, and they have the same Wilson coefficients as \Eq{eq:WilsonForR} after removing the tildes from the soft mass parameters.  In addition, there are eight $R$-symmetry-violating operators at dimension 5 and 6 which contribute to bino LOSP decay.  Their associated Wilson coefficients can again be found by matching.\footnote{There are also analogous results in the case of an uneaten goldstino in the absence of an $R$-symmetry, as long as tildes are added to the soft mass parameters and $\widetilde{G}_L$ is replaced with $\zeta$.  See \App{app:rviolatingdecays}.}
\begin{align}
\mathcal{O}^5_{\sslash{R},B} & =  C^5_{\sslash{R},B} \frac{M_1}{F} i \lambda \sigma^{\mu \nu} \widetilde{G}_L F_{\mu \nu}, \label{eq:op5-B} & C^5_{\sslash{R},B} & = \frac{1}{\sqrt{2}}, \\
\mathcal{O}^5_{\sslash{R},H_u \cdot H_d} & =   C^5_{\sslash{R},H_u\cdot H_d} \frac{ \mu}{F} \lambda \widetilde{G}_L (H_u \cdot H_d), \label{eq:op5-ud} & C^5_{\sslash{R},H_u \cdot H_d} & = 0,\\
\mathcal{O}^5_{\sslash{R},H_u} & =   C^5_{\sslash{R},H_u} \frac{ \mu}{F} \lambda \widetilde{G}_L H_u^\dagger H_u, \label{eq:op5-u} & C^5_{\sslash{R},H_u}& = \frac{g'}{\sqrt{2}} \left(\frac{B_\mu}{\mu^2} - \frac{M_1}{2 \mu}\right), \\
\mathcal{O}^5_{\sslash{R},H_d} & =   C^5_{\sslash{R},d} \frac{ \mu}{F}  \lambda \widetilde{G}_L H_d^\dagger H_d, \label{eq:op5-d} &C^5_{\sslash{R},H_d}& = -\frac{g'}{\sqrt{2}} \left(\frac{B_\mu}{\mu^2} - \frac{M_1}{2 \mu}\right), \\
\mathcal{O}^6_{\sslash{R},1} & = \frac{ C^6_{\sslash{R},1}}{F} i \widetilde{G}_L^\dagger \bar{\sigma}^\mu \lambda (H_u \cdot D_\mu H_d)^*, & C^6_{\sslash{R},1} & = - \frac{g' B_\mu }{\sqrt{2} \mu^2},\\
\mathcal{O}^6_{\sslash{R},2} & =  \frac{ C^6_{\sslash{R},2}}{F} i \widetilde{G}_L^\dagger \bar{\sigma}^\mu  \lambda (D_\mu H_u \cdot H_d)^*, & C^6_{\sslash{R},2} & = \frac{g' B_\mu}{\sqrt{2} \mu ^2}, \\
\mathcal{O}^6_{\sslash{R},3} & = \frac{ C^6_{\sslash{R},3}}{F} i \widetilde{G}_L^\dagger \bar{\sigma}^\mu \lambda (H_u \cdot D_\mu H_d), & C^6_{\sslash{R},3} & = 0, \\
\mathcal{O}^6_{\sslash{R},4} & =  \frac{ C^6_{\sslash{R},4}}{F} i \widetilde{G}_L^\dagger \bar{\sigma}^\mu  \lambda (D_\mu H_u \cdot H_d), & C^6_{\sslash{R},4} & = 0. \label{eq:op6-4}
\end{align}
The first operator $\mathcal{O}^5_B$ is exactly the second term in \Eq{Lgrav}.  The terms proportional to $M_1$ in $C^5_{\sslash{R},H_u}$ and $C^5_{\sslash{R},H_d}$ derive from the third term in \Eq{Lgrav}, which contains the auxiliary field $D$.  The remaining contributions arise from the $R$-symmetry-violating $B_\mu$ term.

Looking at these Wilson coefficients, one might (erroneously) conclude that in the higgsino decoupling limit, a bino LOSP should dominantly decay to a gravitino via a physical higgs instead of via a $\gamma/Z$.  After all, the leading order bino-goldstino-higgs couplings come from four dimension-5 operators---$\mathcal{O}^5_R$, $\mathcal{O}^5_{\sslash{R},H_u \cdot H_d}$, $\mathcal{O}^5_{\sslash{R},H_u}$, and $\mathcal{O}^5_{\sslash{R}, H_d}$---which are enhanced by a factor of $\mu/m_\lambda$ compared to the bino-goldstino-$\gamma/Z$ coupling from $\mathcal{O}^5_B$.  

However, we know this not to be the case for the gravitino.  From conservation of the supercurrent, the decay rate for $\lambda \rightarrow h^0 + \widetilde{G}_L$ given in \Eq{eq:gravh} is suppressed in the decoupling limit by a factor of $\mathcal{O}(m_\lambda^2 M_Z^2 / \mu^4)$ from the decay rates for $\lambda \rightarrow \gamma/Z + \widetilde{G}_L$ given in \Eqs{eq:gravgamma}{eq:gravZ}.  Apparently, when calculating the decay rate of a bino LOSP to a higgs boson and a gravitino using the decoupling limit effective field theory, the contributions to the amplitude from the dimension 5, dimension 6, and dimension 7 operators yield cancellations up to three orders in the $m_\lambda/\mu$ expansion.

\subsection{Miraculous Cancellations}
\label{subsec:cancellations}

The easiest way to see that there must be a cancellation is to go back to the gravitino coupling from \Eq{Lgrav} before integrating out the higgsino.  We can make a standard SUSY transformation on all of our visible sector fields with infinitesimal parameter $\widetilde{G}_L / F$,
\begin{eqnarray}
\phi & \rightarrow & \phi + \frac{1}{F} \psi \widetilde{G}_L, \label{gravredef}
\end{eqnarray}
with similar expressions for other fields.  This is an allowed field redefinition since it leaves the one-particle states unchanged.  Since the coefficients of the SUSY-breaking mass terms and the couplings of $\widetilde{G}_L$ are identical up to a sign, the coupling terms (at order $1/F$) \emph{cancel} under this transformation.  This cancellation is special to the eaten goldstino and does not in general occur for an uneaten goldstino.  The SUSY-respecting part of the lagrangian will clearly remain unchanged under this field redefinition except for terms proportional to $\partial_\mu \widetilde{G}_L$.  Thus, $\widetilde{G}_L$ only couples derivatively to MSSM particles, and does so in exactly the manner described by the supercurrent formalism of \Eq{eq:supercurrent}.  

It is also instructive to see how this cancellation works in the decoupling limit effective field theory.  The $\lambda \rightarrow h^0 + \widetilde{G}_L$ decay may still be completely parametrized as a Yukawa interaction as in \Eq{eq:yukawa} for the leading two orders in $m_\lambda/\mu$:\footnote{The diagrams featuring two dimension 5 operators connected by a virtual bino or wino cancel separately.}
\bea
\mathcal{L} & = & - \frac{M_Z \mu \sin \theta_W}{\sqrt{2} F} \left( C^5_{\textrm{net}} + \frac{ m_\lambda}{\mu} C^6_{\textrm{net}} \right) \lambda \widetilde{G}_L h^0. \label{eq:yukawa-bis}
\eea
The coefficients $C^5_{\textrm{net}}$ and $C^6_{\textrm{net}}$ have new contributions proportional to $B_\mu$ and $M_1$:
\begin{eqnarray}
C^5_{\textrm{net}} & = & \frac{(m_{H_u}^2 - m_{H_d}^2 ) \cos (\alpha + \beta) - 2 B_\mu \sin( \alpha + \beta)}{\mu^2} + \frac{M_1}{\mu} \sin (\alpha + \beta), \label{eq:c5net-grav} \\
C^6_{\textrm{net}} & = & \frac{ m_{H_d}^2 \cos \beta \sin \alpha + m_{H_u}^2 \sin \beta \cos \alpha - B_\mu \cos (\beta - \alpha)}{\mu^2}. \label{eq:c6net-grav}
\end{eqnarray}
If one uses the tree-level relations for the parameters in the higgs potential (see \App{app:treehiggs}), these simplify considerably:
\be
C^5_{\textrm{net}} = - \frac{M_1 \cos 2 \beta}{\mu} + \mathcal{O}\left(\frac{M_Z^2}{\mu^2}\right), \qquad
C^6_{\textrm{net}} = \cos 2 \beta +  \mathcal{O}\left(\frac{M_Z^2}{\mu^2}\right).
\ee
We see that the $\mathcal{O}(1)$ term in $C^5_{\textrm{net}}$ have cancelled entirely, and the $\mathcal{O}(M_1 / \mu)$ term, which arose from the $\lambda \widetilde{G}_L D$ term in \Eq{Lgold}, cancels against $C^6_\textrm{net}$ since $M_1 = m_\lambda$ at this order.  Diagrammatically, the first cancellation is among the diagrams in \Fig{fig:dim5decay}, and the second cancellation is among those in \Fig{fig:dim6cancel}.  There is yet another cancellation at the next order in $\mu$ involving dimension 7 operators, but it is not instructive to show it explicitly here; it may be verified using the methods of \App{sec:allorders} after diagonalizing the neutralino mass matrix order by order in $\mu$.

\FIGURE[t]{
\includegraphics[scale=0.19]{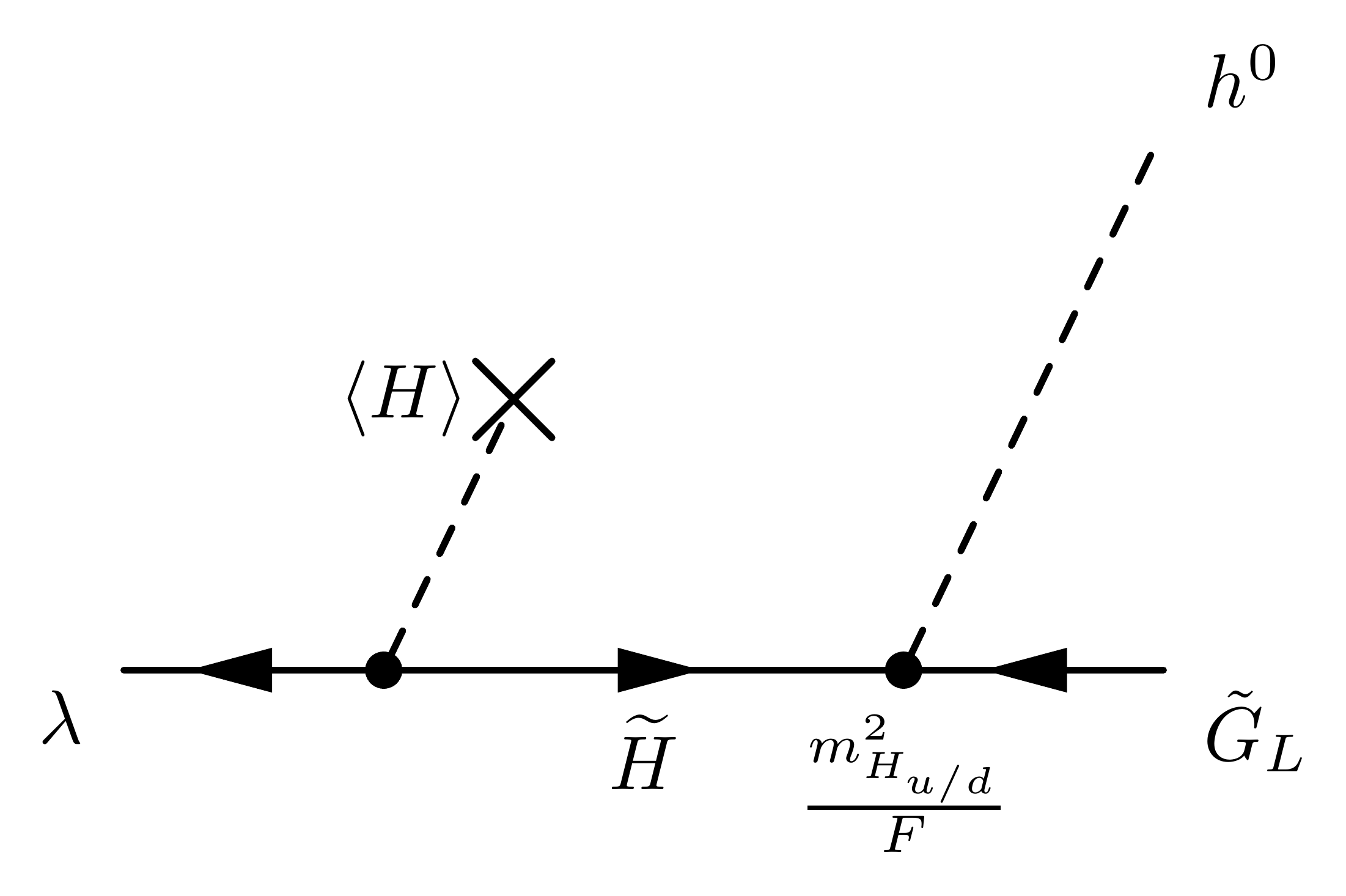} $\qquad$
\includegraphics[scale=0.19]{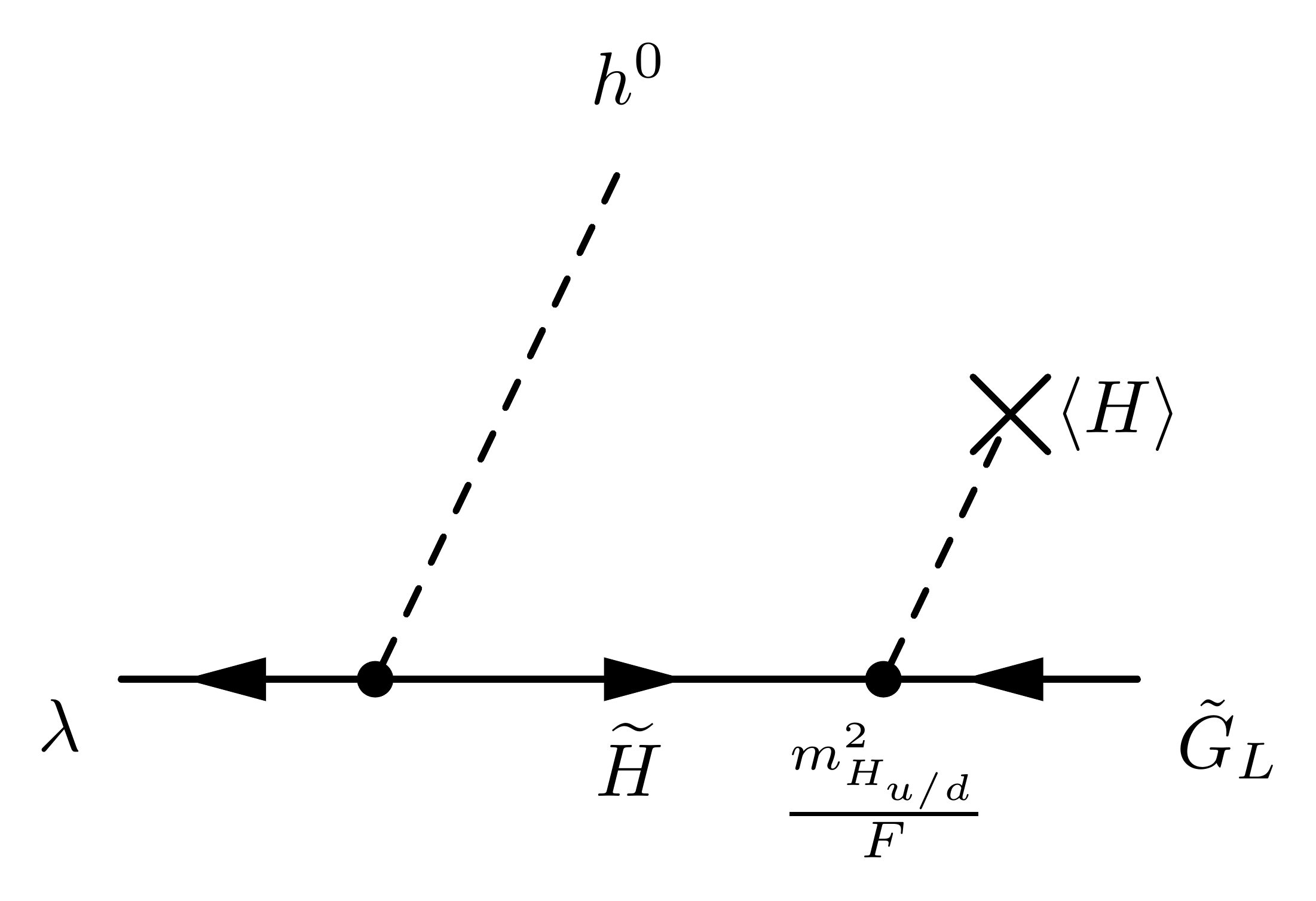} $\qquad$
\includegraphics[scale=0.19]{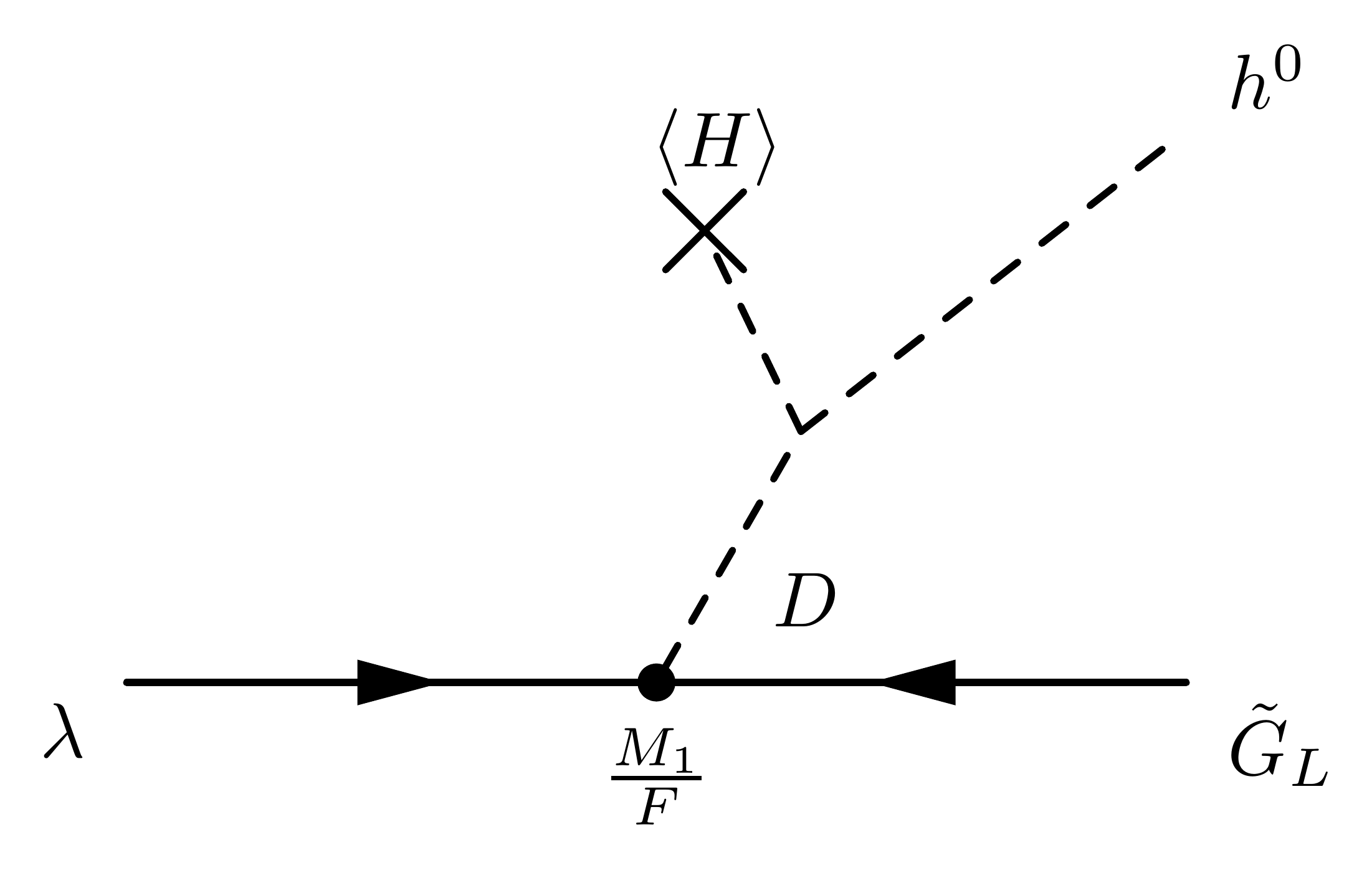}
\caption{These diagrams, which we would expect to yield $\mathcal{O}(\mu^0)$ contributions to the $\lambda \rightarrow h^0 + \widetilde{G}_L$ amplitude, cancel among themselves.}
\label{fig:dim6cancel}
}

\subsection{Why Goldstini are Different}
\label{subsec:goldstinidifference}

These miraculous cancellations for the gravitino case, removing the leading three orders of contributions to the bino LOSP decay to higgs, are very specific to the gravitino and the values of its associated Wilson coefficients.  There is much more freedom in choosing the couplings of the uneaten goldstino.  Concretely, the Wilson coefficients feature the \emph{tilded} versions of soft SUSY-breaking mass parameters, recalling $\widetilde{M}_i = M_{i,2} \cot \theta - M_{i,1} \tan \theta$ from \Eq{eq:tilde}.  These tilded parameters need not satisfy any a priori relation among themselves, and thus the cancellations above will not occur in general for a goldstino.  

Said another way, the mechanisms which ensured the cancellations for the gravitino are not applicable in the goldstino case.    The field redefinition of \Eq{gravredef} made it manifest that the gravitino couples derivatively to observable sector fields, but the same cannot be done in general for the uneaten goldstino.  We could attempt to remove one such coupling with the same sort of SUSY transformation, with
\be
\phi \rightarrow \phi + \frac{1}{F} \frac{\widetilde{m}_\phi^2}{m_\phi^2} \psi \zeta, \label{eq:fieldredef-gold}
\ee
but unless $\widetilde{m}_\phi^2/m_\phi^2 = \widetilde{M}_a / M_a$ for all scalar and gaugino mass terms, there is no transformation that will remove all such couplings and make $\zeta$ purely derivatively coupled.

Thus, one expects a variety of counterintuitive LOSP decay patterns in the presence of goldstini, such as wrong-helicity decays like $\tilde{\tau}_R \rightarrow \tau_L + \zeta$, flavor-violating decays, or reshuffled neutralino/chargino branching fractions.  Of course, the phenomenological differences between a longitudinal gravtino and an uneaten goldstino are highlighted when the ``standard'' decay is forbidden.   This is precisely the case for our bino LOSP in the higgsino decoupling and $R$-symmetric limit, where the standard $\gamma/Z$ decay is suppressed and the novel $h^0$ mode can dominate.

\section{Branching Ratio Results}
\label{sec:results}

We now discuss the bino LOSP branching ratios in the presence of multiple SUSY breaking sectors, using the $R$-symmetric setup from \Fig{fig:Rsetup}.  In the bulk of parameter space, the decay mode $\lambda \rightarrow \psi\bar{\psi} + \zeta$ is suppressed, so we will first focus on the branching ratios to higgs and $Z$ bosons, neglecting any three-body decays.  A brief discussion of what happens away from the $R$-symmetric limit appears in \Sec{sec:Rvioldecays}.

\subsection{Higgs and $Z$ Boson Branching Ratios}

When three-body decays can be neglected, the dominant phenomenology is determined by the two parameters
\be
\epsilon \equiv \frac{m_\lambda \tan \beta}{\mu}, \qquad \tan \gamma  \equiv  \frac{\widetilde{m}^2_{H_u}}{\widetilde{m}^2_{H_d}},
\ee
previously mentioned in \Eqs{eq:xidef}{eq:gammadef}.  Using the partial widths calculated in \Eq{eq:decayrate-higgs} and \Eq{eq:decayrate-Z}, the branching ratio for the bino LOSP decay to $h^0$ or $Z$, assuming both are kinematically allowed, may be expressed in the relatively compact form:
\be
\textrm{Br}(\lambda \rightarrow h^0 \zeta) =  \frac{\left(\frac{\epsilon^{-1} - \epsilon_0^{-1}}{\omega}\right)^2}{1+ \left(\frac{\epsilon^{-1} - \epsilon_0^{-1}}{\omega}\right)^2}, \qquad \textrm{Br}(\lambda \rightarrow Z \zeta) =  \frac{1}{1+ \left(\frac{\epsilon^{-1} - \epsilon_0^{-1}}{\omega}\right)^2}.
\ee
In particular, the branching ratio to $Z$ bosons is a Lorentzian in $\epsilon^{-1}$ and is thus negligible for small $\epsilon$, as expected.  The Lorentzian is centered at $\epsilon^{-1}_0$ with a width $\omega$,
\begin{align}
\epsilon^{-1}_0 &=  \frac{1 - \tan \gamma \tan^2 \beta}{2 \tan^2 \beta (\tan \gamma - 1)}, \\
\omega &=  \frac{1 + \tan \gamma \tan^2 \beta}{2 \tan^2 \beta (\tan \gamma - 1)} \left(\frac{m_\lambda^2 - M_Z^2}{m_\lambda^2 - m_{h^0}^2} \sqrt{1 + 2 \frac{M_Z^2}{m_\lambda^2}} \right),
\end{align}
where the precise values of $\epsilon^{-1}_0$ and $\omega$ depend on the higgs soft mass ratio $\tan \gamma$, $\tan \beta$, and various kinematic factors.  Of course, additional three-body decays, whether to fermions or to multiple higgs or Z bosons, will spoil the simplicity of these expressions.

\FIGURE[t]{
\includegraphics[scale=0.65]{figures/contour}
\includegraphics[scale=0.65]{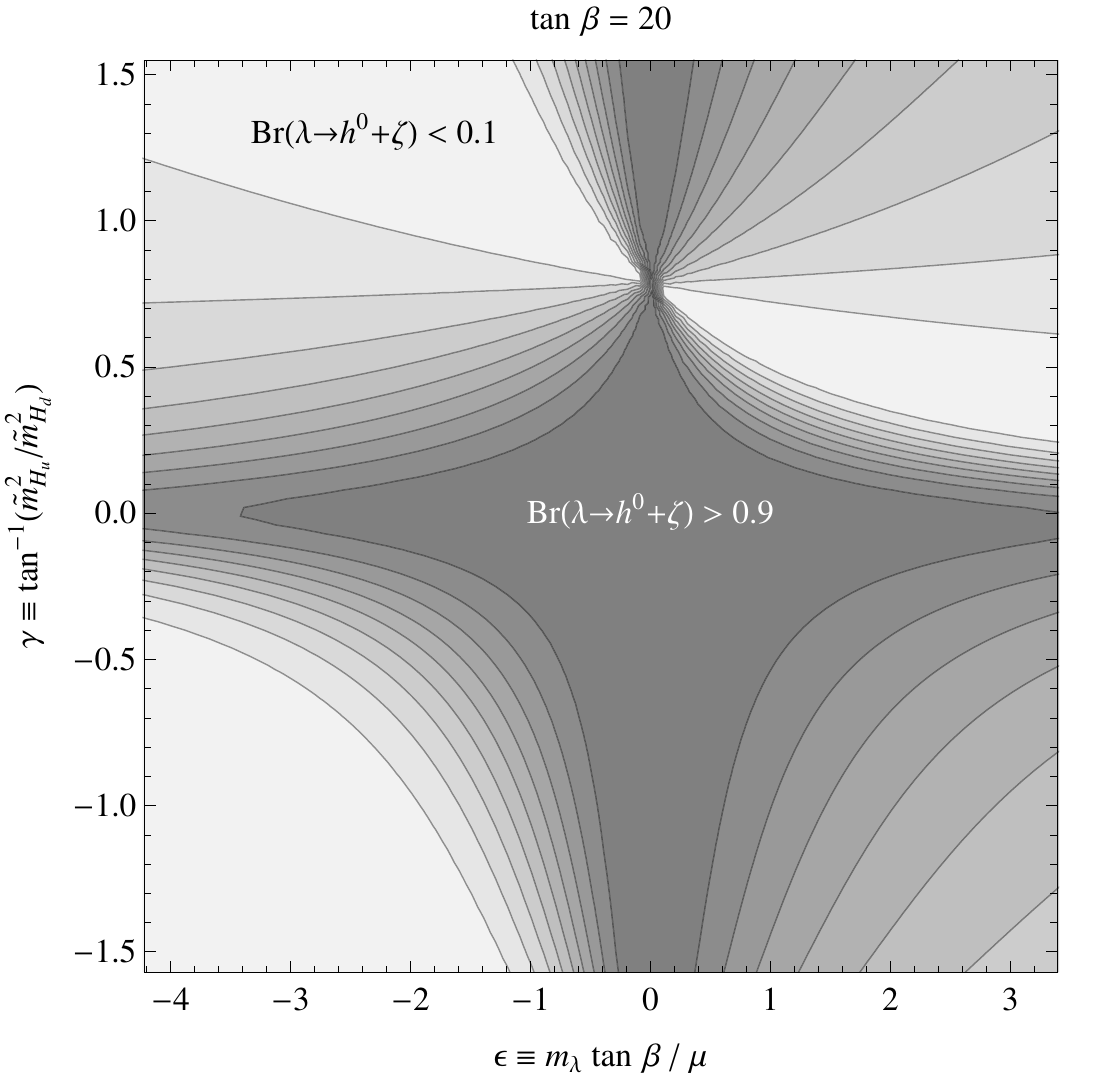}
\caption{Branching ratios for $\lambda \rightarrow h^0 + \zeta$ in the $\epsilon$--$\gamma$ plane for $\tan \beta = 5$ (left, same as \Fig{fig:xiplot}) and $\tan \beta = 20$ (right), respectively.  The remaining branching ratio is dominated by $\lambda \to Z + \zeta$.  The main differences between the two plots arise because at larger $\tan \beta$, the kinematically excluded region $m_\lambda < m_h^0$ (which bounds the left plot) is not encountered until larger $\epsilon$.  In this and the remaining plots, we have fixed $M_1 = 155~\GeV$ and $m_{h^0} = 120~\GeV$, which are mainly relevant for setting the phase space factors in the partial widths.}
\label{fig:epsgamma}
}

Plots of the branching ratio to higgs in the $\epsilon$--$\gamma$ plane are shown in \Fig{fig:epsgamma}, and slices through that plane are shown in \Figs{fig:epsplots}{fig:gammaplots}.  In the latter plots, the solid lines are the all-orders tree-level calculations from \App{sec:allorders}, while the dashed lines are the analytic results obtained using the higgsino decoupling effective theory from \Sec{sec:eft} (while still using the all-orders result for the physical LOSP mass $m_\lambda$).

\FIGURE[t]{
$\begin{array}{c c}
\includegraphics[scale=0.65]{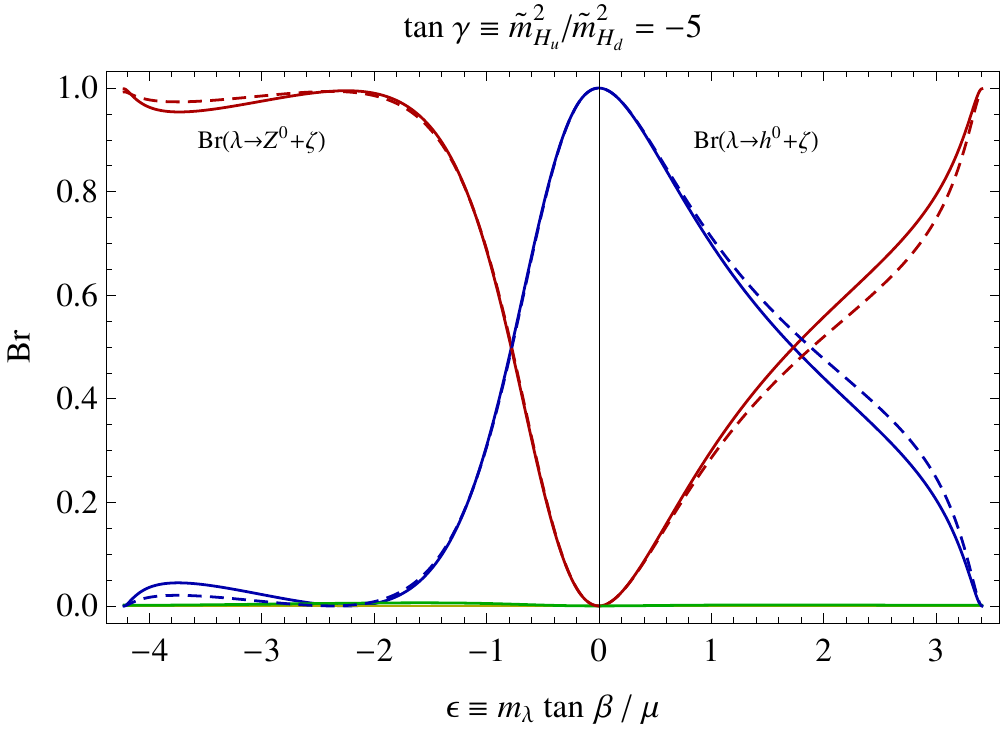} \quad & \quad
\includegraphics[scale=0.65]{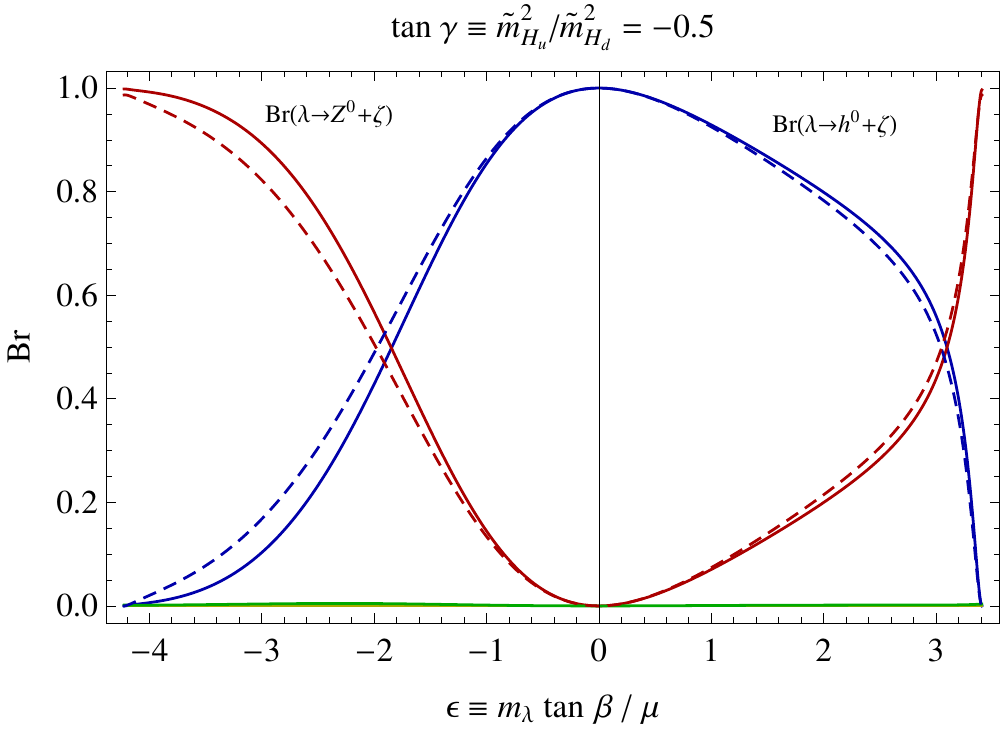} \\
\includegraphics[scale=0.65]{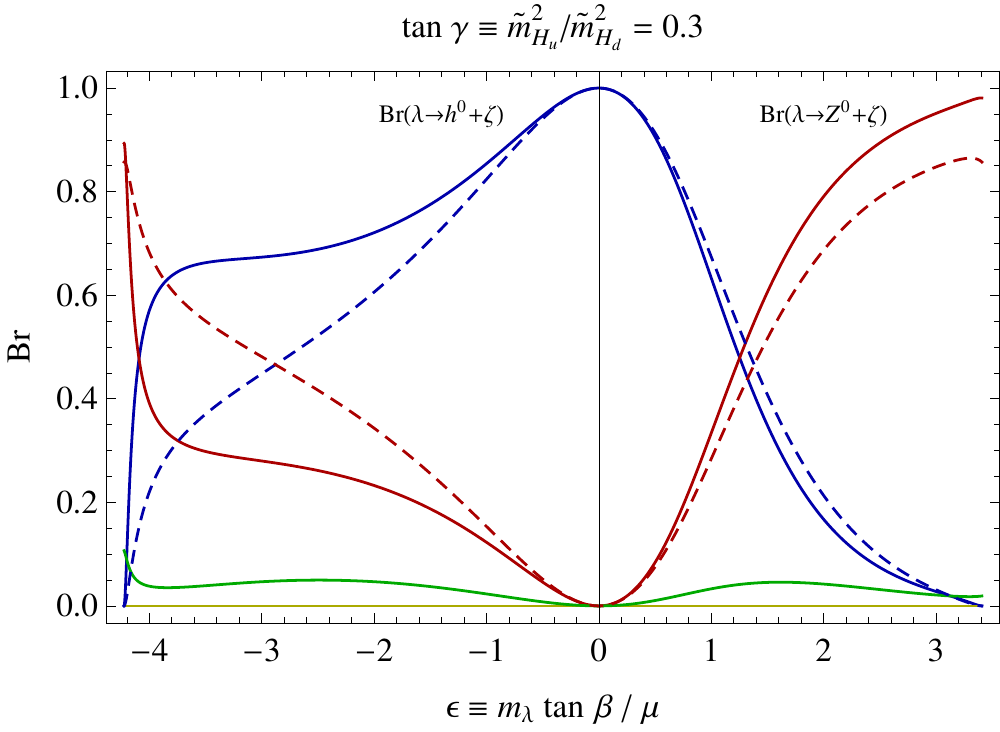} \quad & \quad
\includegraphics[scale=0.65]{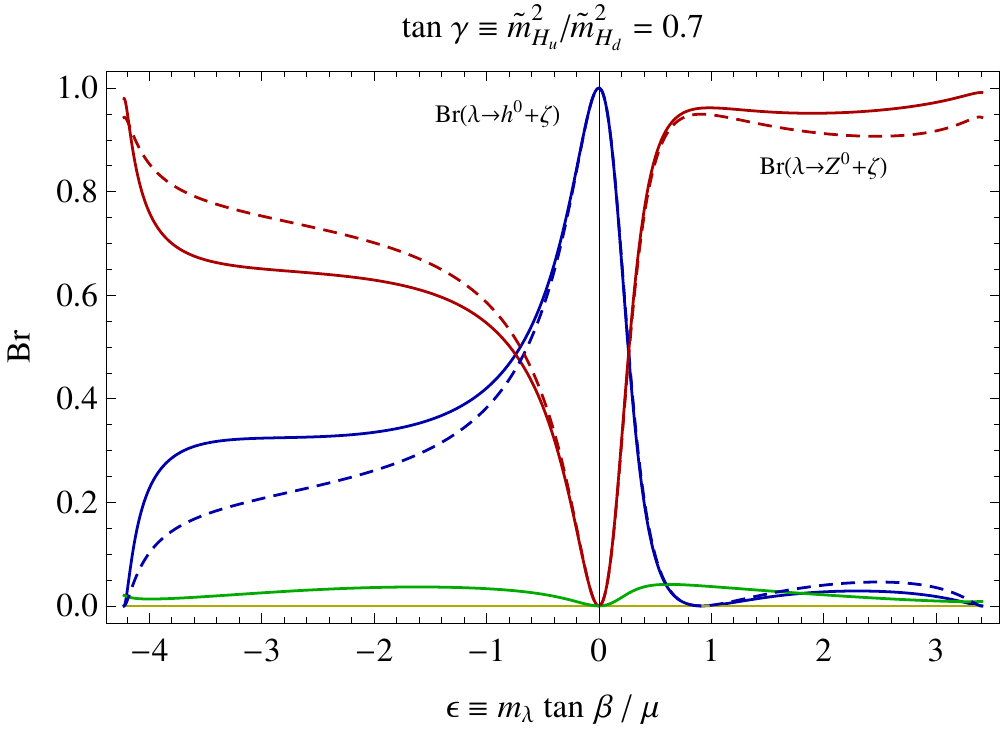}
\end{array}$
\caption{Branching ratios for the bino LOSP as a function of $\epsilon$ for fixed values of $\tan \gamma$.  These are all slices of the left plot in \Fig{fig:epsgamma} with $\tan \beta = 5$, $M_1 = 155~\GeV$, and $m_{h^0} = 120~\GeV$. The solid curves are the all-orders result from \App{sec:allorders}, while the dashed curves are from the higgsino decoupling effective theory in \Sec{sec:eft}.  The curves are $\Br(\lambda \rightarrow h^0 \zeta)$ (blue), $\Br(\lambda \rightarrow Z \zeta)$ (purple), $\Br(\lambda \rightarrow \psi \bar{\psi} \zeta)$ (green), and $\Br(\lambda \rightarrow \gamma \zeta)$ (yellow).  The decay to higgses dominates in the small $\epsilon$ limit, with the next most relevant mode being the $Z$.  The branching ratio to difermions is calculated using the results of \Sec{sec:difermions}, taking the parameter $\rho$ defined in \Eq{eq:rho} to be $1.0$.  As advertised, this branching ratio to difermions is very suppressed, and the branching ratio to photons is essentially zero.}
\label{fig:epsplots}
}

\FIGURE[t]{
$\begin{array}{c c}
\includegraphics[scale=0.65]{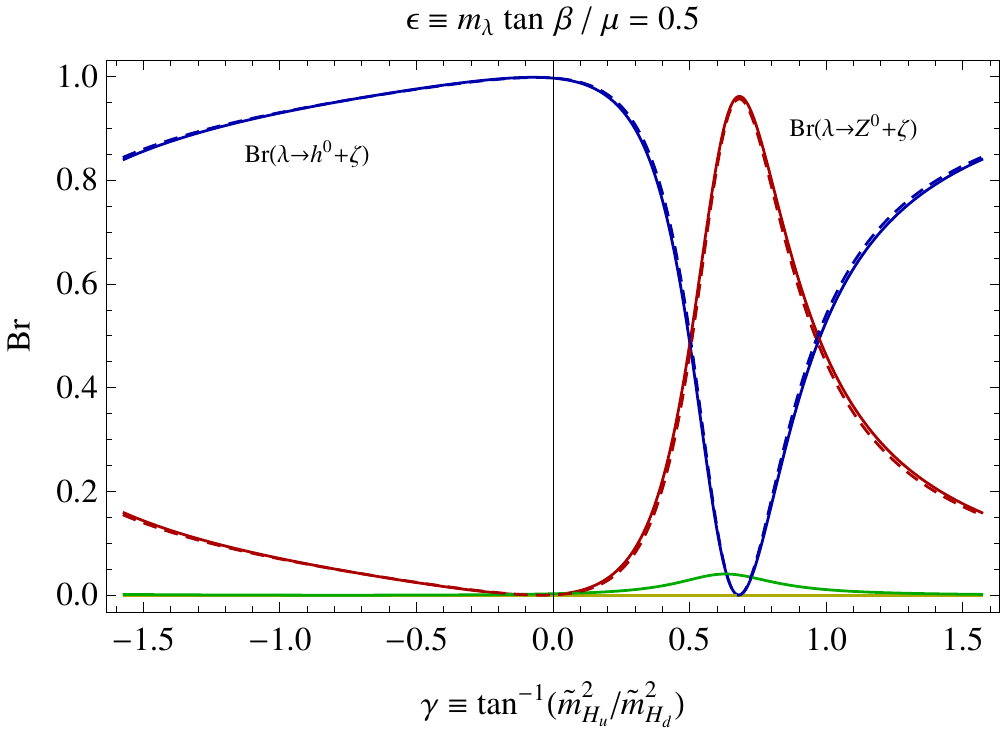} \quad & \quad 
\includegraphics[scale=0.65]{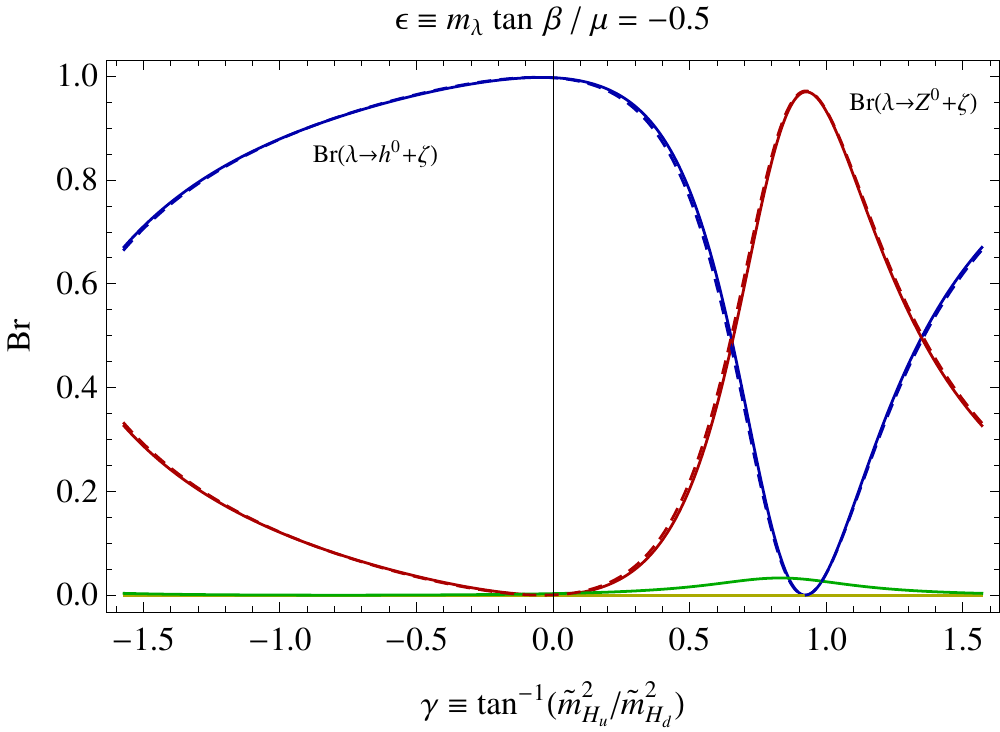} \\
\includegraphics[scale=0.65]{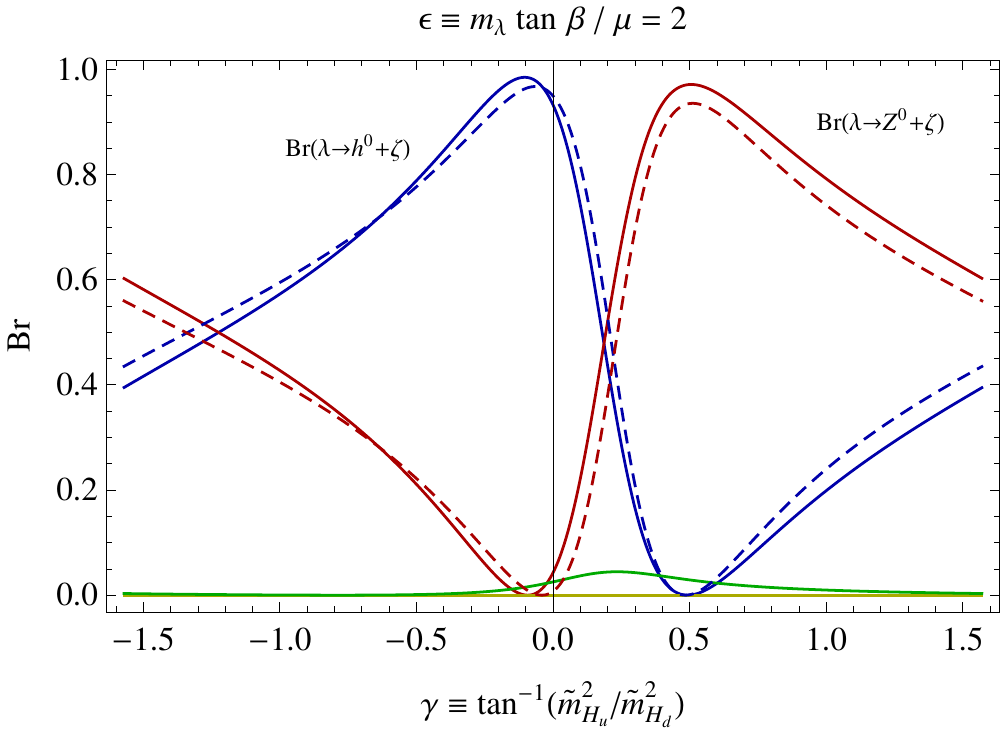} \quad & \quad
\includegraphics[scale=0.65]{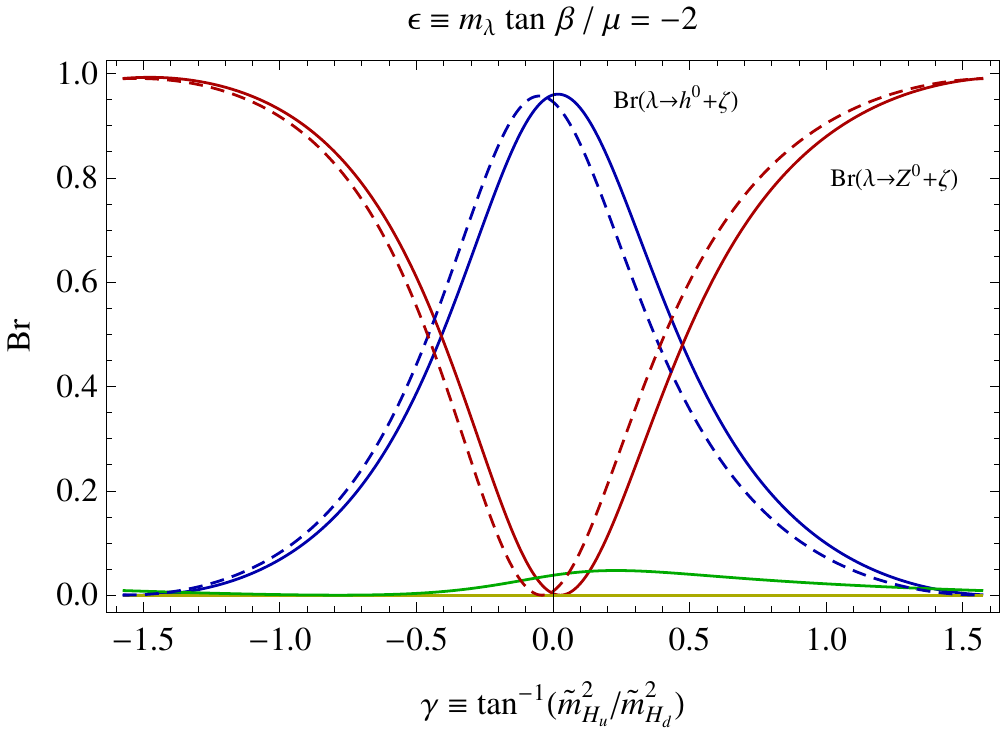}
\end{array}$
\caption{Same as \Fig{fig:epsplots}, but with branching ratios given as a function of $\gamma$ for fixed values of $\epsilon$.}
\label{fig:gammaplots}
}

The small $\epsilon$ limit corresponds to the extreme higgsino decoupling regime, where not only $|\mu| \gg m_\lambda$, but the $\tan \beta$ suppressed dimension 5 operator $\mathcal{O}^5_R$ dominates over the dimension 6 operators.  Thus, generically, for small $\epsilon$, the decay is overwhelmingly to higgs bosons, as expected.  However, there is an exception for the region around $\tan \gamma = 1$.  When $\tan \gamma = 1$, $\widetilde{m}_{H_u}^2 - \widetilde{m}_{H_d}^2$ and $C^5_{\textrm{net}}$ are both zero and the branching ratios to higgs and $Z$ bosons should be roughly equal up to phase space factors.  For $\tan \gamma$ slightly removed from unity (downwards for $\epsilon > 0$, upwards for $\epsilon < 0$), $C^5_{\textrm{net}}$ will destructively interfere with $C^6_{\textrm{net}}$ and the $Z$ mode will dominate. 

Moving away from small $\epsilon$, we expect the $Z$ branching ratio to increase, as the contributions from dimension 6 operators to bino decay are roughly equal for the higgs and $Z$ modes.  This is shown in \Fig{fig:epsplots}.  The effects of interference between the dimension 5 and dimension 6 operators on the higgs amplitude also grow more pronounced for larger $\epsilon$.  For $\epsilon > 0$, the interference is destructive for $\tan \gamma \in (1/\tan^2 \beta, 1)$, and vice versa for $\epsilon < 0$.\footnote{The operative relative sign is that between $\mu$ and $m_\lambda$.  The $\mathcal{O}^5$ operator features an odd power of $\mu$, while the $m_\lambda$ factor comes from the $C^6$ operators, whose only non-vanishing contributions feature the Dirac equation applied to the external bino spinor.}

For extremely large $\epsilon$, the approximations based on being in the higgsino decoupling limit break down as $m_\lambda / \mu$ approaches $\mathcal{O}(1)$. Ultimately, the higgs mode is kinematically excluded once the mass of the lightest neutralino (by now predominantly higgsino) drops below the higgs mass.

\subsection{Difermion Branching Ratio}
\label{sec:difermions}

In most of parameter space, the decay mode $\lambda \rightarrow \psi\bar{\psi} + \zeta$ is suppressed.  We can see this most clearly by comparing the decay rate to all fermion species to the decay rate to a $Z$:
\begin{eqnarray}
\frac{\sum_\psi \Gamma_{\psi \bar{\psi}}}{\Gamma_Z} & = & \frac{\alpha_{\textrm{EM}}}{3 \pi \sin^2 2 \theta_W} \frac{\sum_i Y_i^2 \tau_i^2 f_i}{\left(C^6_{\textrm{net},Z}\right)^2} \frac{m_\lambda^2}{M_Z^2} \left(1 - \frac{M_Z^2}{m_\lambda^2}\right)^{-2} \left(1 + 2 \frac{M_Z^2}{m_\lambda^2} \right)^{-1},
\end{eqnarray}
where $\tau_i \equiv \widetilde{m}_{\phi_i}^2 / m_{\phi_i}^2$ and $f_i \equiv f[m_{\phi_i}^2/m_\lambda^2]$, with the function $f$ defined in \Eq{eq:lightscalar}.

For concreteness, consider the limit where $\tan \beta \gg 1$, $|\mu|, m_{\phi_i} \gg m_\lambda$, and the $\tau_i$ are all equal to a common value $\tau_0$.  The sum over SM fermion hypercharges (excluding the presumably kinematically inaccessible top) is 103/12. Assuming that the tree-level relations between the soft masses approximately hold, $C^6_{\textrm{net},Z} = \tau_0$.  All the $\tau_i$ values then cancel, and the net result is
\begin{eqnarray}
\frac{\sum_\psi \Gamma_{\psi \bar{\psi}}}{\Gamma_Z} & \approx & \frac{1}{107} \frac{m_\lambda^2}{M_Z^2}  \left(1 - \frac{M_Z^2}{m_\lambda^2}\right)^{-2} \left(1 + 2 \frac{M_Z^2}{m_\lambda^2} \right)^{-1}.
\end{eqnarray}
This ratio obtains a minimum of around $1/28$ at $m_\lambda \approx 140$ GeV, and it is smaller than $1/10$ for $m_\lambda$ in the approximate range 100--300 GeV. 

Of course, there is one somewhat contrived region of parameter space for which the decay to fermions can dominate; if the sector containing the uneaten goldstino gives no contribution to any of the higgs soft masses, then $\widetilde{m}_{H_u}^2$ and $\widetilde{m}_{H_d}^2$ vanish and the decay via an off-shell sfermion are the only ones allowed.  \Fig{fig:mavgplot} shows that the decay to fermions can indeed dominate if the parameter
\be
\rho \equiv \frac{\widetilde{m}^2_{H_u} + \widetilde{m}^2_{H_d}}{2 \mu^2  \cot \theta} \frac{\sum_i Y_i^2}{\sum_i Y_i^2 \tau_i f_i},
\label{eq:rho}
\ee
with sums taken over all appropriate sfermion species, is tuned close enough to zero.

\FIGURE[t]{
$\qquad$$\qquad$$\qquad$ \includegraphics[scale=0.65]{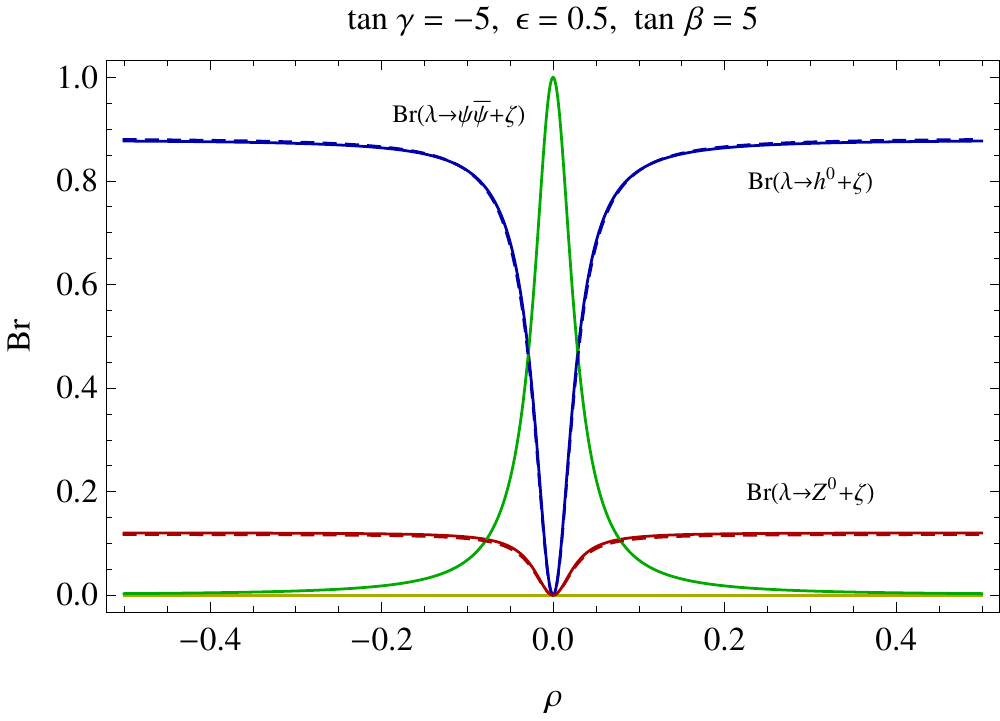} $\qquad$$\qquad$$\qquad$
\caption{Branching ratios for the bino LOSP as a function of the parameter $\rho$ defined in \Eq{eq:rho} below, measuring in effect the relative contributions to the higgs and sfermion mass terms by sector 2.  If this parameter is tuned close to zero, then the higgs and $Z$ modes shut off, leaving only the difermion channel.  For larger values of $\rho$, the difermion channel is suppressed; this occurs generically when the tilded higgs soft mass parameters scale with $\mu^2$, as mentioned in \Sec{sec:decouplingandRlimit}.  For concreteness, all prior figures have used $\rho = 1.0$.}
\label{fig:mavgplot}
}

\subsection{The $R$-violating Regime}
\label{sec:Rvioldecays}

Though not the focus of this paper, we wish to briefly comment on possible $R$-violating decays, for which calculations are given in \App{app:rviolatingdecays}.   As we move away from the $R$-symmetric limit, the LOSP decay to photons is now allowed at tree level, and will generally garner a branching ratio that is at least of the same order as of those to higgs or $Z$.  In \Fig{fig:deltaplot}, we show branching ratios as a function of a parameter $\delta$ which measures the amount of deviation from the $R$-symmetric limit,
\be
\delta \equiv \frac{2}{3} \frac{\tau_1 + \tau_2 + \tau_{B_\mu}}{\tau_{H_u} + \tau_{H_d}}, \label{eq:defdelta}
\ee
with $\tau_i \equiv \widetilde{M}_i / M_i$ for any soft mass(-squared) parameter $M_i$.  In \Fig{fig:deltaplot}, we hold $ \tau_1 = \tau_2 = \tau_{B_\mu}$ and $\tau_{H_d} = \tau_{H_u} = \tau_{\phi_i}$ for simplicity.  When $\delta = 0$ we have the exact $R$-symmetric limit; when $\delta = 1$ we have the ``aligned'' limit in which the uneaten goldstino couples simply as a rescaled version of the gravitino (i.e.\ there is a basis, obtained by making the field redefinition \Eq{eq:fieldredef-gold}, in which it couples only derivatively).  Note in the latter limit the higgs branching ratio effectively shuts off, as expected.

\FIGURE[t]{
$\qquad$$\qquad$$\qquad$ \includegraphics[scale=0.65]{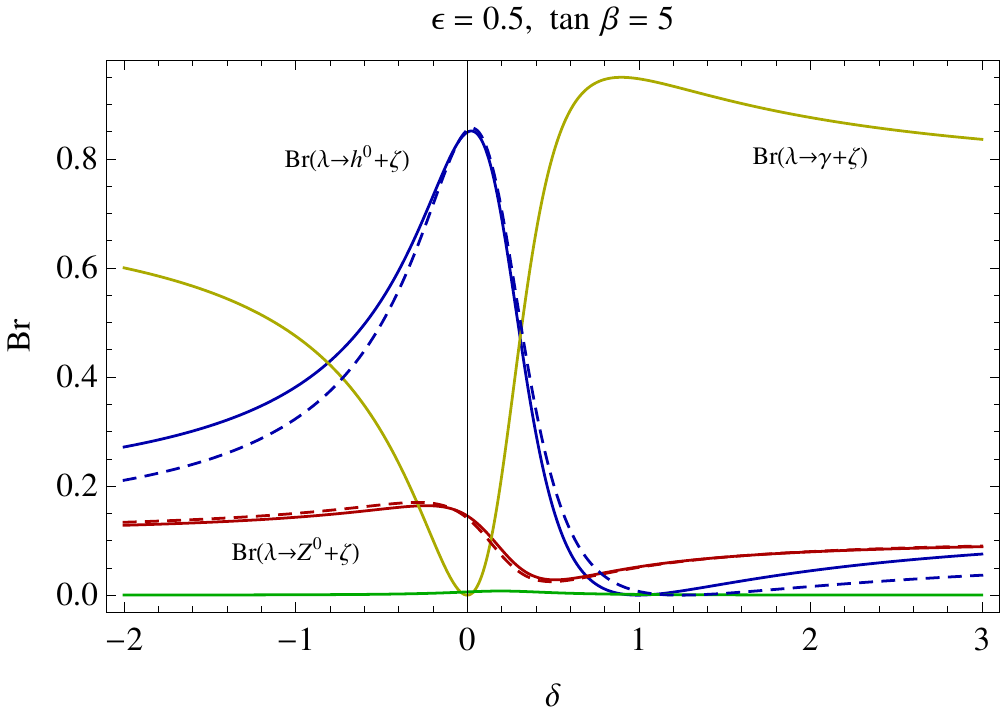} $\qquad$$\qquad$$\qquad$
\caption{Branching ratios for the bino LOSP as a function of the parameter $\delta$, defined in \Eq{eq:defdelta}, that measures the deviation from the $R$-symmetric limit.  When $\delta = 0$, we are in the $R$-symmetric limit of the previous figures.   When $\delta = 1$, the branching ratios for $\lambda \rightarrow X +\zeta$ are exactly what one would predict for $\lambda \rightarrow X + \widetilde{G}_L$ in the more conventional model with only one hidden sector; the photon mode dominates and the higgs mode is highly suppressed.}
\label{fig:deltaplot}
}

The diversity of possible LOSP decay branching ratios shown in \Fig{fig:deltaplot} is reminiscent of mixed neutralino LOSP scenarios, where the LOSP has comparable bino, wino, and higgsino fractions.  Here, however, we are still working in the higgsino decoupling limit, so the interesting pattern of LOSP widths come not from varying the identity of the LOSP but rather from varying how the hidden sectors couple to the SSM.

\section{Conclusion}
\label{sec:conclude}

SUSY breaking scenarios with a light gravitino offer fascinating phenomenological possibilities.  With the LOSP no longer stable, gravitinos could comprise part or all of the dark matter of the universe, and collider experiments could discover extended SUSY cascade decays.  However, the gravitino need not be the only SUSY state lighter than the LOSP.   In the context of multiple SUSY breaking, there is a corresponding multiplicity of goldstini whose masses are all typically proportional to $m_{3/2}$ (or loop suppressed compared SSM soft masses).  Thus, the LOSP may dominantly decay to an uneaten goldstino instead of the gravitino.  Since the couplings of the uneaten goldstino are unconstrained by supercurrent conservation, the LOSP can exhibit counterintuitive decay patterns.

In this paper, we have focused on the case of a bino-like LOSP which decays dominantly to higgs bosons despite having negligible higgsino fraction.  This effect is particularly pronounced in the presence of a $U(1)_R$ symmetry, which suppresses the expected $\lambda\rightarrow \gamma + \zeta$ decay.  By studying which low energy effective operators are generated in the higgsino decoupling limit, we have understood why the mode $\lambda\rightarrow h^0 + \zeta$ dominates in the limit of small $(m_\lambda \tan \beta) / \mu$, and also why there is a non-standard $\lambda\rightarrow Z + \zeta$ decay mode away from that limit.  We have seen explicitly that there are delicate cancellations in the decay width of the LOSP to a gravitino, and the counterintuitive decays of a LOSP to an uneaten goldstino arise from incomplete cancellations.

Similar counterintuitive decay patterns would be present for a wino-like LOSP, and in general, one should contemplate the possibility of any LOSP decay pattern consistent with SM charges.  Those LOSP decays might involve an uneaten goldstino as in this paper, but could also be present with a light axino 
\cite{Kim:1983ia,Rajagopal:1990yx} or a new light hidden sector \cite{ArkaniHamed:2008qp,DeSimone:2010tr,Cheung:2010jx}.  To our mind, the most intriguing possibilities involve copious higgs boson production in the final stages of a SUSY cascade decay, which may offer new higgs discovery modes and give further motivation for boosted higgs searches.  Studying these phenomena is particularly relevant given the expected LHC sensitivity to SUSY scenarios in the coming two years.

\begin{acknowledgments}
J.T. and Z.T. are supported by the U.S. Department of Energy (D.O.E.) under cooperative research agreement DE-FG02-05ER-41360. 
\end{acknowledgments}

\appendix
\section{Tree-Level Higgs Potential}
\label{app:treehiggs}

The MSSM tree-level higgs potential for the neutral higgs sector arises from a combination of $F$-terms, $D$-terms, and three soft SUSY-breaking terms:
\bea
V(H_u^0,H_d^0) & = & (|\mu|^2 + m_{H_u}^2 )|H_u^0|^2 + (|\mu|^2 + m_{H_d}^2) |H_d^0|^2 + B_\mu (H_u^0 H_d^0 + H_u^{0*} H_d^{0 *})\nonumber \\
& &~ + \frac{g^2 + g'^2}{8} (|H_u^0|^2 - |H_d^0|^2)^2.
\eea
Once we recall that 
\bea
M_Z^2 & = & \frac{1}{2} (g^2 + g'^2) \left(\left<H_u^0\right>^2 + \left<H_d^0\right>^2\right), \\
\tan \beta & \equiv & \left<H_u^0\right> / \left<H_d^0\right>,
\eea
we can use the fact that the vacuum must minimize the higgs potential to find relations among these parameters.  
\bea
0 & = & m_{H_u}^2 + |\mu|^2 + B_\mu \cot \beta - M_Z^2 \frac{\cos 2 \beta}{2}, \label{eq:relation-Hu}\\
0 & = & m_{H_d}^2 + |\mu|^2 + B_\mu \tan \beta + M_Z^2 \frac{\cos 2 \beta}{2}. \label{eq:relation-Hd}
\eea
It is convenient to take linear combinations of these relations, one without $|\mu|^2$ and one without $B_\mu$:
\bea
0 & = & (m_{H_u}^2 - m_{H_d}^2) \sin 2 \beta + 2 B_\mu \cos 2 \beta - M_Z^2 \frac{\sin 4 \beta}{2}, \label{eq:relation-Bmu} \\
0 & = & m_{H_u}^2 \sin^2 \beta - m_{H_d}^2 \cos^2 \beta - |\mu|^2 \cos 2 \beta - M_Z^2 \frac{\cos 2 \beta}{2}. \label{eq:relation-mu}
\eea
In the higgsino decoupling limit ($|\mu|^2, m_{A^0}^2 \gg M_Z^2$), we may neglect the terms proportional to $M_Z^2$.  Also in the same limit, the tree-level relation for the physical higgs mixing angle $\alpha$ simplifies considerably: 
\be
\tan 2 \alpha = \tan 2 \beta \, \frac{m_{A^0}^2 + M_Z^2}{m_{A^0}^2 - M_Z^2} \quad \Rightarrow \quad \alpha = \beta - \pi/2 + \mathcal{O}\left(\frac{M_Z^2}{m_{A^0}^2}\right). \label{eq:alpha}
\ee

Once one applies \Eq{eq:alpha}, the relations \Eqs{eq:relation-Bmu}{eq:relation-mu} are precisely those which cause the cancellation of the $\lambda \rightarrow h^0 + \widetilde{G}_L$ amplitude at the first two orders in $\mu/M_1$ in \Eqs{eq:c5net-grav}{eq:c6net-grav}.
Another linear combination of \Eqs{eq:relation-Hu}{eq:relation-Hd} gives a (non-independent) relationship that can be useful for simplifying $C^6_{\textrm{net},Z}$,
\be
0 = |\mu|^2 + m_{H_u}^2 \sin^2 \beta + m_{H_d}^2 \cos^2 \beta + B_\mu \sin 2 \beta +  M_Z^2 \frac{\cos^2 2 \beta}{2}. \label{eq:relation-Z}
\ee
A third relation, involving the pseudoscalar mass $m_{A^0}^2$, allows us to solve for all three soft mass parameters if desired:
\bea
B_\mu & = & - \frac{1}{2} m_{A^0}^2 \sin 2 \beta, \label{eq:Bmu}\\
m_{H_u}^2 & = & -|\mu|^2 + m_{A^0}^2 \cos^2 \beta + M_Z^2 \frac{\cos 2 \beta}{2}, \\
m_{H_d}^2 & = & -|\mu|^2 + m_{A^0}^2 \sin^2 \beta - M_Z^2 \frac{\cos 2 \beta}{2}.
\eea
Of course, all of the above relations are valid only at tree-level, and one does expect corrections to these relations from the same loop effects needed to raise the physical higgs mass above the LEP bounds.

\section{$\boldsymbol{R}$-Symmetry Violating Decays}
\label{app:rviolatingdecays}

In the body of this paper, we focused on the setup in \Fig{fig:Rsetup} where sector 2 preserves an $R$-symmetry.  If the sector 2 does not preserve an $R$-symmetry, then there are many more allowed operators that can mediate the decay of a bino LOSP to the uneaten goldstino.  They are exactly those previously given for the gravitino in \Sec{subsec:additionaloperators}, except with the replacement of $\widetilde{G}_L$ with $\zeta$ and with all soft masses tilded. 

A decay to photon at tree-level is now allowed through the usual operator 
\be
\mathcal{O}^5_{\sslash{R},B} =  \frac{i \widetilde{M}_1}{\sqrt{2} F} \lambda \sigma^{\mu \nu} \zeta F_{\mu \nu}, \label{eq:op5-B-bis}
\ee
with resultant decay rate
\be
\Gamma_\gamma = \frac{\widetilde{M}_1^2 m_\lambda^3 \cos^2 \theta_W}{16 \pi F^2}.
\ee

The couplings of the bino LOSP to the physical higgs $h^0$ and any further couplings to the $Z$ not already found in $\mathcal{O}^5_{\sslash{R},B}$ may be parametrized at the first two orders in $m_\lambda / \mu$ as
\be
\mathcal{L} = - \frac{M_Z \mu \sin \theta_W}{\sqrt{2} F} \left[ \left( C^5_{\textrm{net}} + \frac{m_\lambda}{\mu} C^6_{\textrm{net}}\right) \lambda \zeta h^0 - \frac{M_Z}{\mu} C^6_{\textrm{net},Z} \zeta^\dagger \bar{\sigma}^\mu \lambda Z_\mu\right],
\ee
with $C_{\rm net}$ representing the following linear combinations of Wilson coefficients:
\bea
\frac{g'}{\sqrt{2}} C^5_{\textrm{net}} &  = &  \left(C^5_R + C^5_{\sslash{R},H_u \cdot H_d}\right) \cos (\alpha + \beta) \nonumber \\
& & ~ - 2 C^5_{\sslash{R},H_u} \sin \beta \cos \alpha + 2 C^5_{\sslash{R},H_d} \cos \beta \sin \alpha , \\
\frac{g'}{\sqrt{2}} C^6_{\textrm{net}} & = &\left(C^6_{H_u,1} + C^6_{H_u,2}\right) \sin \beta \cos \alpha - \left(C^6_{H_d,1} + C^6_{H_d,2}\right) \cos \beta \sin \alpha \nonumber\\
& & ~ + \left(C^6_{\sslash{R},1}+C^6_{\sslash{R},3}\right) \sin \beta \sin \alpha - \left(C^6_{\sslash{R},2} + C^6_{\sslash{R},4}\right) \cos \beta \cos \alpha, \\
\frac{g'}{\sqrt{2}} C^6_{\textrm{net},Z}  & = & - \left(C^6_{H_u,1} - C^6_{H_u,2}\right) \sin^2 \beta + \left(C^6_{H_d,1} - C^6_{H_d,2}\right) \cos^2 \beta \nonumber \\
& & ~ +  \frac{1}{2} \left(C^6_{\sslash{R},1} - C^6_{\sslash{R},2} - C^6_{\sslash{R},3} + C^6_{\sslash{R},4}\right) \sin 2 \beta.
\eea
Here, the factors of $g'/\sqrt{2}$ are inserted purely for convenience.  In the $R$-symmetric limit, all the $C_{\sslash{R}}$ are of course zero.

For the decay to higgs, the formula \Eq{eq:decayrate-higgs} for the decay rate still holds, but $C^5_{\textrm{net}}$ now has contributions proportional to $\widetilde{B}_\mu$ and $\widetilde{M}_1$, as it did in \Eq{eq:c5net-grav}:\footnote{Again, we use the approximation $\alpha \approx \beta - \pi/2$ from \Eq{eq:alpha}, which is appropriate at this order in $m_\lambda / \mu$.  This eliminates a term proportional to $\widetilde{B}_\mu \cos (\beta - \alpha)$ in $C^6_{\textrm{net}}$.}
\be
C^5_{\textrm{net}} = \frac{(\widetilde{m}_{H_u}^2 - \widetilde{m}_{H_d}^2 ) \sin 2\beta + 2 \widetilde{B}_\mu \cos 2 \beta}{\mu^2} - \frac{\widetilde{M}_1}{\mu} \cos 2 \beta.
\ee
For the decay to $Z$, $C^6_{\textrm{net},Z}$ obtains a term proportional to $\widetilde{B}_\mu$
\be
C^6_{\textrm{net},Z} = - \frac{\widetilde{m}_{H_u}^2 \sin^2 \beta + \widetilde{m}_{H_d}^2 \cos^2 \beta + \widetilde{B}_\mu \sin 2 \beta}{\mu^2},
\ee
and we must also include the effects of $\widetilde{M}_1$ from $\mathcal{O}^5_{\sslash{R},B}$ to find the full decay rate:
\begin{eqnarray}
\Gamma_Z & = & \frac{m_{\lambda}^3 \widetilde{M}_1^2 \sin^2 \theta_W}{16 \pi F^2} \left(1 - \frac{M_Z^2}{m_\lambda^2}\right)^2 \nonumber \\
& &  \times \left(1 + \frac{1}{2} \frac{M_Z^2}{m_\lambda^2} - \frac{3 M_Z^2 C^6_{\textrm{net},Z}}{\widetilde{M}_1 m_\lambda}  + \left(\frac{M_Z C^6_{\textrm{net},Z}}{\sqrt{2} \widetilde{M}_1}\right)^2 \left(1 + 2 \frac{M_Z^2}{m_\lambda^2}\right)\right). \label{eq:fullZrate}
\end{eqnarray}

For the gravitino, $C^6_{\textrm{net},Z}$ simplifies to unity at this order due to the tree-order relation \Eq{eq:relation-Z}, and the complicated expression in \Eq{eq:fullZrate} simplifies to the same result we obtained from the supercurrent in \Eq{eq:gravZ}, as it must.  We demonstrated in \Sec{sec:gravitinocase} that the decay rate to higgs bosons simplifies similarly and in fact completely cancels at this order.  For an uneaten goldstino, however, such cancellations do not generically occur, unless the ratio $\tau_i \equiv \widetilde{M}_i/ M_i$ is equal for all soft SUSY-breaking mass(-squared) terms $M_i$.  It is precisely when all the $\tau_i$ are equal that one can make the field redefinition \Eq{eq:fieldredef-gold} to make the goldstino couple only derivatively to visible-sector fields.  In this limit, it would couple in exactly the same way as the longitudinal gravitino, except with an enhancement  factor of $\tau^2 \sim \cot^2 \theta$.  Of course, we should not expect such alignment to occur in general (if only due to loop corrections), so a generic uneaten goldstino will have branching ratios to photons, $Z$s, and higgses of roughly the same order of magnitude, as suggested by \Fig{fig:deltaplot}.

\section{All-Orders Tree-Level Calculation}
\label{sec:allorders}

The higgsino decoupling limit studied in \Sec{sec:eft} is convenient for understanding the physical origin of the counterintuitive LOSP decays, but it is tedious in practice for moderate values of $\mu$.  Instead of integrating out the higgsinos and finding an arbitrarily long series of operators and associated Wilson coefficients, we may conduct the calculation with the original lagrangian in the mass eigenstate basis.  As long as one can explicitly diagonalize the $4\times 4$ neutralino mass matrix (analytically or numerically), one can perform the full tree-level calculation to all orders in $\mu$.  

To do so, we parametrize the relevant interactions from \Eq{eq:fulllagrangian} as follows:
\bea
\mathcal{L}  &= & - \frac{1}{2} M_{i j} \chi_i \chi_j + \rho_i \zeta \chi_i  - \frac{1}{2} Y_{i j} \chi_i \chi_j h^0 + y_i \zeta \chi_i h^0 \nonumber \\
& &~ +  G_{i j} \chi_i^\dagger \bar{\sigma}^\mu \chi_j Z_\mu + L_i \, i \zeta \sigma^{\mu \nu} \chi_i \partial_\mu Z_\nu .
\eea
In the $\{\lambda_B$,$\lambda_3$,$\widetilde{H}_d^0$,$\widetilde{H}_u^0 \}$ basis, the neutralino mass matrix is \cite{Martin:1997ns}\be
M = \left( \begin{array}{c c c c} 
M_1 & 0 & - M_Z c_\beta s_W & M_Z s_\beta s_W \\
0 & M_2 & M_Z c_\beta c_W & - M_Z s_\beta c_W \\
- M_Z c_\beta s_W & M_Z c_\beta c_W & 0 & -\mu \\
M_Z s_\beta s_W & - M_Z s_\beta c_W & - \mu & 0 
\end{array}\right), \ee
the linear mixing with the uneaten goldstino is
\be
\rho  = \frac{v}{\sqrt{2} F} \left(\begin{array}{c} \frac{1}{4}  g' v \widetilde{M}_1 \cos 2 \beta \\ - \frac{1}{4}  g v \widetilde{M}_2 \cos 2 \beta \\ \widetilde{m}_{H_d}^2 c_\beta + \widetilde{B}_\mu s_\beta \\ \widetilde{m}_{H_u}^2 s_\beta + \widetilde{B}_\mu c_\beta \end{array} \right), \\
\ee
the couplings to the physical higgs boson are
\be
Y = \frac{1}{2} \left(
\begin{array}{c c c c}
0 & 0 & g' s_\alpha & g' c_\alpha \\
0 & 0 & - g s_\alpha & - g c_\alpha \\
g' s_\alpha & - g s_\alpha & 0 & 0 \\
g' c_\alpha & -g c_\alpha & 0 & 0
\end{array}\right), 
\qquad y = \frac{1}{\sqrt{2} F}  \left( \begin{array}{c} - \widetilde{M}_1 M_Z s_W \sin(\alpha + \beta) \\ \widetilde{M}_2 M_Z c_W \sin(\alpha + \beta) \\ \widetilde{B}_\mu c_\alpha - \widetilde{m}_{H_d}^2 s_\alpha \\ \widetilde{m}_{H_u}^2 c_\alpha - \widetilde{B}_\mu s_\alpha \end{array} \right),
\ee
and the couplings to the $Z$ boson are
\be
G = \frac{g}{2 c_W}
\left(\begin{array}{c c c c}
0 & 0 & 0 & 0 \\
0 & 0 & 0 & 0 \\
0 & 0 & 1 & 0 \\
0 & 0 & 0 & -1
\end{array}\right), \qquad
L = \frac{\sqrt{2}}{F}\left(\begin{array}{c}\widetilde{M}_1 s_W \\ - \widetilde{M}_2 c_W \\ 0 \\ 0 \end{array}\right).
\ee
In the above matrixes, we have used the notation $c_\theta \equiv \cos \theta$ and $s_\theta \equiv \sin \theta$, with $W$ standing for the weak mixing angle $\theta_W$.  

To calculate the decays of the lightest neutralino, we go to the mass eigenstate basis:
\be
M \rightarrow M' =  P^T M P,
\ee
with $P$ chosen to make $M'$ diagonal.  Note that we treat the linear mixing with the uneaten goldstino as an insertion, which is valid to leading order in $1/F$.  The other matrices and vectors rotate as
\be
\rho \rightarrow \rho' = P^T \rho, \qquad Y \rightarrow Y' =  P^T Y P,
\ee
and so forth.  The full tree-level amplitude for the decay of the lightest neutralino to a higgs/$Z$ and a goldstino is thus:
\bea
\Gamma_{h ^0}& = & \frac{m_\lambda}{16 \pi} \left(y'_1 - \sum_i \frac{Y'_{1 i} \rho'_i}{m_{\chi^0_i}}\right)^2 \left(1 - \frac{m_{h^0}^2}{m_\lambda^2}\right)^2, \\
\Gamma_Z & = & \frac{m_\lambda^3}{16 \pi}   \left(1 - \frac{M_Z^2}{m_\lambda^2}\right)^2 \left( \frac{L_1^{'2}}{2} \left(1 + \frac{1}{2} \frac{M_Z^2}{m_\lambda^2}\right) - 3 \frac{L'_1 K'}{ m_\lambda} +  \frac{K'^2}{M_Z^2} \left(1 + 2 \frac{M_Z^2}{m_\lambda^2}\right) \right),
\eea
where the neutralino masses are labeled by $m_{\chi^0_i}$, the LOSP mass is $m_\lambda \equiv m_{\chi^0_1}$, and
\be
K' \equiv \sum_i \frac{G'_{1 i} \rho'_i}{m_{\chi^0_i}}. \\
\ee

A similar calculation for difermion production is beyond the scope of this work; it would in general need to include the effects of $A$-terms, finite fermion masses, and sfermion mixing for the third generation, as well as possible interference from difermions produced by off-shell $Z$ bosons.

\bibliography{GoldstiniHiggs}

\end{document}